# Epitaxial growth and semiconductor properties of $NiGa_2O_4$ spinel for $Ga_2O_3$/NiO interfaces

Kingsley Egbo[1], Emily M. Garrity[2], Shivashree Shivamade Gowda[3], Saman Zare[3], Ethan A. Scott[3], Glenn Teeter[1], Brooks Tellekamp[1], Vladan Stevanovic[2], Patrick E. Hopkins[3], Andriy Zakutayev,[1†] Nancy Haegel[1]

[1] National Laboratory of the Rockies, Golden, CO 80401, USA
[2] Colorado School of Mines, Golden, CO 80401, USA
[3] University of Virginia, Charlottesville, VA 22904, USA
[†]Andriy.Zakutayev@nrel.gov

## ABSTRACT

Unintentionally formed interfacial layers are ubiquitous in semiconductor devices that operate at extreme conditions. However, these layers' structure and properties often remain unknown due to the thinness of these naturally formed interphases. Here, we report on the intentional epitaxial growth and semiconductor properties of $NiGa_2O_4$ spinel layers that form at $Ga_2O_3$/NiO interfaces used in high-power and high-temperature electronic devices. Cubic spinel $NiGa_2O_4$ films of 10-50 nm thicknesses and low surface roughness (~ 2 nm) were grown using pulsed laser deposition at a substrate temperatures in the 300-900 °C range on $\alpha$-$Al_2O_3$ and $\beta$-$Ga_2O_3$ substrates of different orientation. The optical absorption onset (3.6-3.9 eV) and thermal conductivity (4-9 W $m^{-1}$ $K^{-1}$) vary systematically with substrate temperature, consistent with theoretical predictions of varying Ni and Ga cation ordering on the spinel lattice. The valence band offset between $NiGa_2O_4$ and $\beta$-$Ga_2O_3$ is determined to be 1.8 eV. The $NiGa_2O_4$-based p-n heterojunction devices on $Ga_2O_3$ (001) substrates exhibit a rectification ratio of $10^8$ (for $\pm$2V) and a turn-on voltage of 1.4 V, maintaining diode behavior up to 600 ºC. These results highlight the potential of $NiGa_2O_4$ as a $p$-type interlayer in $Ga_2O_3$-based devices and shows a new approach to investigate such interfacial layers.

## INTRODUCTION

Monoclinic $\beta$-$Ga_2O_3$ is an emerging ultrawide bandgap (UWBG) semiconducting oxide, recognized for its promise in electronic and photonic applications. $\beta$-$Ga_2O_3$ is predicted to sustain a critical electric field strength of 8 MV/cm, potentially surpassing the performance of current commercial materials such as SiC and GaN in high-voltage devices and power electronics[1]. With a bandgap of 4.8 eV, $\beta$-$Ga_2O_3$ is also suited for solar-blind detection, sensing, and transparent conductor applications within the deep ultraviolet (UV) spectrum[2]. Over the past decade, significant advancements have been made in the growth of large-diameter, high-quality $\beta$-$Ga_2O_3$ crystals and wafers[3,4], alongside improvements in thin-film epitaxy, leading to the development of high-performance electronic and optical devices. These combined advances in quality $\beta$-$Ga_2O_3$ manufacturing suggest commercial and performance advantages in power electronics, as indicated by techno-economic analyses and cost modeling[5,6].

However, progress in power electronic devices utilizing oxide-based UWBG materials like $\beta$-$Ga_2O_3$ has been constrained by the challenge of ambipolar doping. Although n-type doping is relatively well established in wide bandgap oxides, the compensating donors and flat valence bands typical of these materials render p-type doping challenging[7]. Additionally, the wide bandgap constrains p-type dopants to primarily act as deep acceptors, further complicating the ability to achieve robust p-type conductivity[8]. Despite significant progress in developing oxide-based materials for various applications, the difficulty in realizing reliable p-type doping remains a substantial bottleneck. To achieve the theoretical performance projections for $\beta$-$Ga_2O_3$ in p-n heterojunction devices and junction field-effect transistors (JFETs), the presence of p-type layers is essential. This challenge of *p*-type doping has limited the possibility of further improving breakdown voltages in β-$Ga_2O_3$ based devices due to inability to achieve charge balance and electric field modulation.

Considering the difficulties associated with p-type doping in $\beta$-$Ga_2O_3$, alternative strategies involving the heteroepitaxial growth of p-type oxide layers have been explored as potential substitutes. These efforts have included the integration of p-type oxides such as NiO[9], SnO[10], and α-$Cr_2O_3$[11] with $\beta$-$Ga_2O_3$. For instance, the use of p-NiO has enabled one group to fabricate vertical power diodes with breakdown voltages over 10 kV[12,13]. Recent work has indicated that the NiO/$Ga_2O_3$ interface can operate at elevated temperatures of up to 400 °C[14], but may become unstable at higher temperature, limiting the application range for NiO/$\beta$-$Ga_2O_3$ based high-temperature and high-power devices. While α-$Cr_2O_3$/$Ga_2O_3$ heterojunctions have demonstrated better interfacial stability up to 600 ºC[15], their lower barrier height results in high leakage currents. Consequently, further research is needed to identify and investigate more compatible and high-temperature stable p-type oxide materials for integration with $\beta$-$Ga_2O_3$ in power devices.

We recently reported the development of spinel-type ternary $NiGa_2O_4$ thin films as alternative *p*-type oxide layers that can be heteroepitaxially integrated into $\beta$-$Ga_2O_3$/NiO interface junctions, forming rectifying contacts useful for device applications[16] . It was shown that $NiGa_2O_4$ forms naturally at the $\beta$-$Ga_2O_3$/NiO interface during operation at 600 ºC. Intentional insertion of the thin $NiGa_2O_4$ layer at this interface during synthesis improved both rectification ratio and breakdown voltage. These beneficial effects were explained by theoretical defect calculations that showed the undoped $NiGa_2O_4$ to be a weakly p-type semiconductor. Other literature suggests that $NiGa_2O_4$ is a promising wide bandgap oxide material with reported *p*-type characteristics: this material has been investigated for applications in batteries[17] , photocatalysis[18], and organic gas sensing[19] . However, its structure and properties as a p-type layer for optoelectronic and power applications, particularly in thin-film semiconductor form, remain underexplored.

In this work, we report on the synthesis of $NiGa_2O_4$ thin films using pulsed laser deposition (PLD) and the characterization of their optoelectronic, thermal, and interfacial properties. The $NiGa_2O_4$ thin films were grown by PLD on α-$Al_2O_3$(00.1) substrates with a 40-50 nm thicknesses, and on β-$Ga_2O_3$ (100), (010) and (001) substrates with 3-8 nm thicknesses. Structural characterization revealed that $NiGa_2O_4$ (111) grows epitaxially on α-$Al_2O_3$ (00.1), $NiGa_2O_4$ (002) grows epitaxially on β-$Ga_2O_3$(100), and $NiGa_2O_4$ (220) stabilizes on β-$Ga_2O_3$ (010). Spectroscopic ellipsometry measurements indicated an optical absorption onset in the range of 3.5 to 3.9 eV for the $NiGa_2O_4$ thin films, and thermal conductivity measured with time-domain thermoreflectance (TDTR) ranges from 4 to 9 W $m^{-1}$ $K^{-1}$, both depending on the film growth temperature (300 °C – 900 °C). Theoretical calculations of the $NiGa_2O_4$ spinel indicated that these variations in optical and thermal properties may be due to varying cation ordering. Furthermore, the valence band offset at the $NiGa_2O_4$/β-$Ga_2O_3$ (100) interface was determined to be 1.8 eV, suggesting that the $NiGa_2O_4$/β-$Ga_2O_3$ heterostructure forms a type II band offset, making it promising for p-n heterojunction applications. Vertical $NiGa_2O_4$/$Ga_2O_3$ p-n diodes, fabricated on a 1 μm thick hydride vapor phase epitaxy (HVPE) β-$Ga_2O_3$ (001) layer, exhibited an excellent rectification ratio of $10^8$ (at ±2 V) and a turn-on voltage of 1.4 V, with the temperature dependence of the diode parameters analyzed from the I–V–T characteristics up to 600 ºC. Our findings indicate that thin films of spinel $NiGa_2O_4$ hold promise as p-type oxide interlayers for β-$Ga_2O_3$/NiO devices in high-power and high-temperature applications.

## RESULTS AND DISCUSSION

$NiGa_2O_4$ thin films were grown on $\alpha$-$Al_2O_3$(00.1) and $\beta$-$Ga_2O_3$ substrates by PLD, with the growth monitored for surface structure and roughness by Reflection High-Energy Electron Diffraction (RHEED). The resulting samples were characterized by X-ray diffraction (XRD) for epitaxial relations, and spectroscopic ellipsometry (SE) for bandgap and refractive index. Thermal conductivities were measured with time-domain thermoreflectance (TDTR), supported by infrared variable-angle spectroscopic ellipsometry (IR-VASE) for phonon lifetime measurements. The valence band and core level spectra were measured by x-ray photoelectron spectroscopy(XPS) on the samples with varying $NiGa_2O_4$ thickness and analyzed for band offsets with $Ga_2O_3$ by fitting (described in the text). Vertical $NiGa_2O_4$/$Ga_2O_3$ (001) heterojunction devices were fabricated with Ti/Au and Ni/Au contacts of different diameters, characterized by temperature-dependent current-voltage (I-V) measurements, and analyzed by thermionic emission and Poole-Frenkel models. Theoretical calculations of structural energies were performed using density functional theory using the GGA+U approximation, and the electronic structure calculations were performed using the HSE06 + $G_0W_0$ approach. More details about the synthesis, characterization, fabrication, and calculation methods are provided in the Methods section.

### Epitaxial growth and optical properties of $NiGa_2O_4$ on $Al_2O_3$

Smooth and uniform growth of epitaxial $NiGa_2O_4$ layers on $Al_2O_3$ (00.1) substrates was confirmed through in-situ reflection high-energy electron diffraction (RHEED) and ex-situ x-ray diffraction (XRD). In Fig. 1a, RHEED images of the bare $Al_2O_3$ and the $NiGa_2O_4$ layers at two different azimuthal angles exhibit streaky patterns, indicating a possible layer-by-layer growth mode of crystalline films with smooth surface morphology. Further evidence for the smoothness of the films is provided by XRD data presented in Fig. 1b, which shows the (222) reflection for $NiGa_2O_4$ across four substrate temperature conditions. The (222) peak shifts to higher angles as the substrate temperature increases, suggesting enhanced compressive strain within the $NiGa_2O_4$ layer due to lattice matching with the underlying $Al_2O_3$ substrate. The $NiGa_2O_4$ hexagonal close-packed spacing of 2.92 Å aligns with the $Al_2O_3$ oxygen sublattice spacing of ~2.74 Å. A representative XRD rocking curve measurement of the $NiGa_2O_4$ (222) peak, FWHM=0.059°, is shown in Fig. S1 of the Supplementary Information. The low FWHM values are indicative of both the high quality of the layer and the high quality of the $Al_2O_3$ substrate crystal, compared to $Ga_2O_3$ substrate crystals discussed later (Fig. S2, Fig. S4, Fig. S5, Table S1).

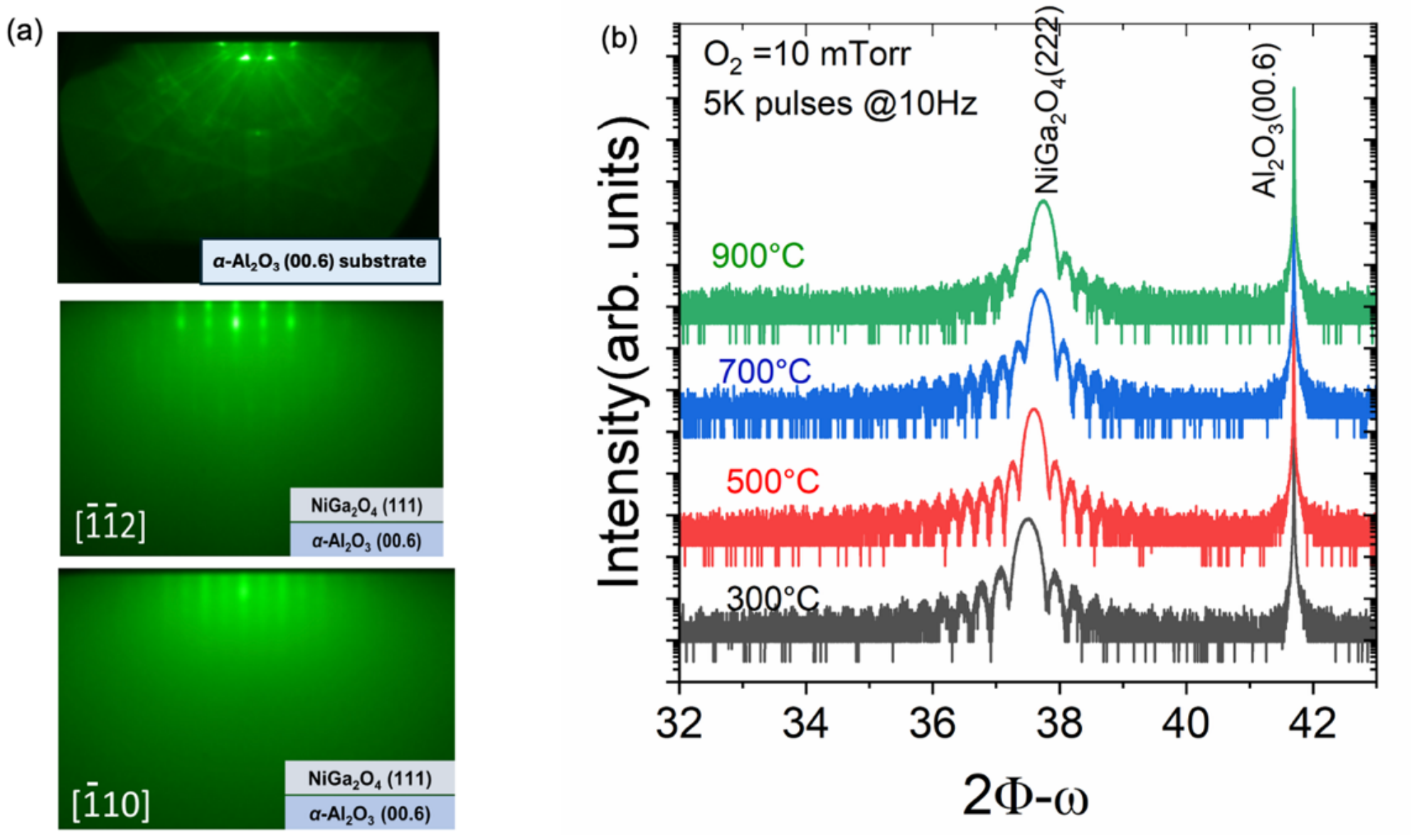

Fig. 1**.** (a) Reflection high energy electron diffraction (RHEED) images of α-$Al_2O_3$(00.1) substrate and the $NiGa_2O_4$(111) thin films grown on c-Al2O3 for [112]-azimuth and [110]-azimuth. (b) Out-of-plane 2θ-ω scan showing heteroepitaxial spinel-type $NiGa_2O_4$(111) on α-$Al_2O_3$(00.1) for thin films deposited at $p(O_2)$ of 10 mTorr and several substrate temperature setpoints as indicated.

The optical absorption onset of $NiGa_2O_4$ increases with substrate temperature during growth. Spectroscopic ellipsometry data, as shown in Fig. 2a, reveal that the absorption onset shifts from 3.6 eV to 3.8 eV as the substrate temperature increases, as determined through Tauc analysis (inset of Fig. 2a). These values are consistent with theoretical bandgap values determined at the GW level for normal, inverse, and disordered spinel crystal structures with different site occupancies on the spinel lattice (Fig. 2b). Variation in the optical absorption edge may be related to cation disorder contributing to the decrease in the bandgap for thin films grown at lower substrate temperatures, as shown in the past for other oxide spinels[20,21], and would be discussed by theoretical calculations presented next. The refractive index (*n = 2.0-2.2*) and extinction coefficient (*k*) spectra and their analysis for $NiGa_2O_4$ are presented in Fig. S6.

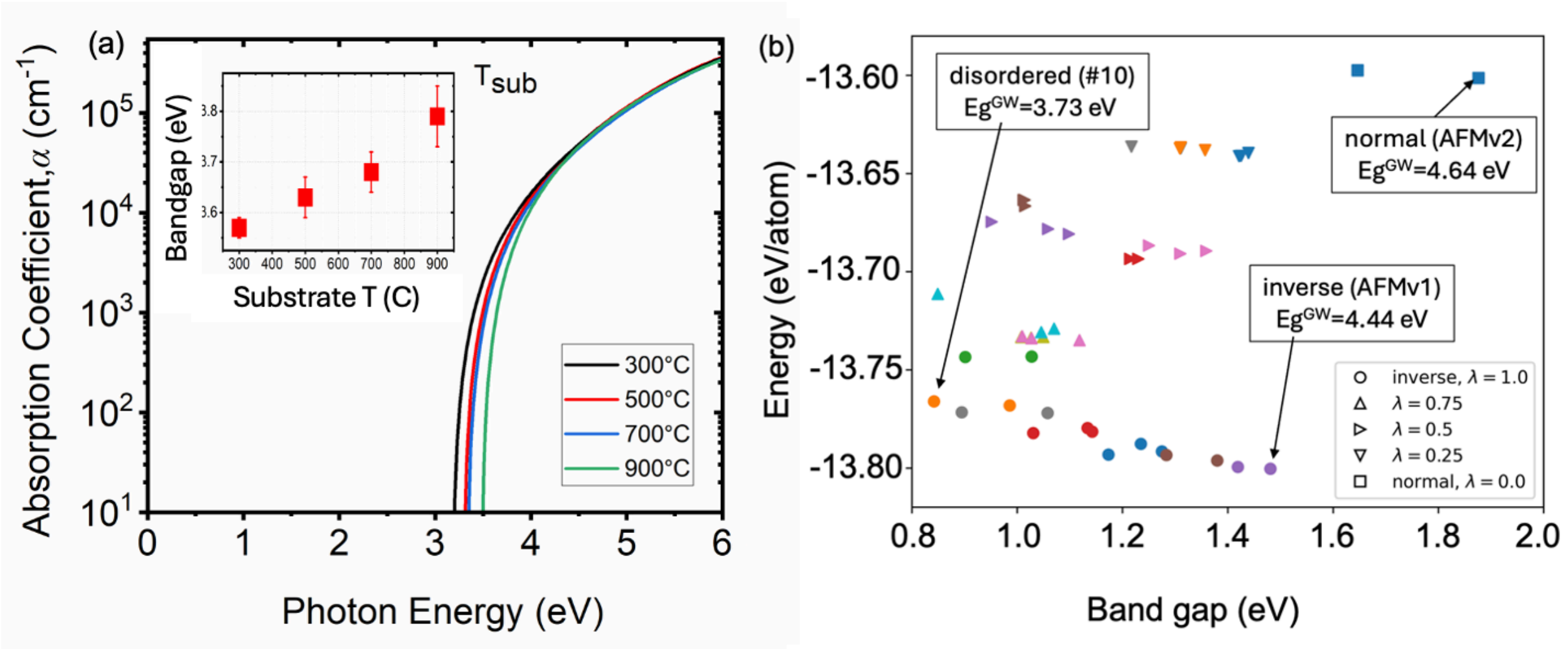

Fig. 2. (a) Absorption coefficient spectra determined from spectroscopic ellipsometry of the $NiGa_2O_4$(111) on α-$Al_2O_3$(00.1) for samples grown by PLD at different substrate temperature setpoints. Inset shows the extrapolated bandgaps in the range of 3.6-3.8 eV for the $NiGa_2O_4$ thin films. (b) Calculated band gaps and total energies for various 28-atom $NiGa_2O_4$ spinel configurations, with colored data points representing different anti-ferromagnetic spin configurations (AFMv#) of Ni for a given cation ordering (marker shape). Band gap decreases with increasing cation disorder.

**Theoretical stability and electronic structure calculations of $NiGa_2O_4$**

To investigate the influence of cation ordering on the band gap and the electronic structure of $NiGa_2O_4$, we used first-principles calculations. The spinel crystal structure of $NiGa_2O_4$ is a cubic lattice of oxygen anions where nickel or gallium cations can occupy either octahedral ($Oh$) or tetrahedral ($Td$) interstitial sites. It is useful to describe the spinel configuration using the dimensionless quantity λ defined in the Methods section. In the ground state (T=0 K), the cations are completely ordered, creating either a fully normal or fully inverse spinel, depending on the elements. In a fully normal spinel (λ=0), all $Td$ sites are occupied by Ni and all $Oh$ sites are occupied by Ga. In a fully-inverse spinel (λ=1), all $Td$ sites are occupied by Ga and the $Oh$ sites are occupied 50%-50% by Ga and Ni. Thus, a fully inverse spinel is a class of configurations rather than a single specific crystal structure. At elevated temperature, cation disordering is possible and the degree of inversion can vary over the range $0 > \lambda > 1$, with maximal configurational entropy occurring at $\lambda = 0.67$ [22].

Previous studies[20,22–24] have demonstrated the impact of cation disorder on changing electronic structure, resulting in lowering of the bandgap. To test this sensitivity, we calculated the electronic structure of various cation configurations of a 28-atom unit cell of $NiGa_2O_4$. In addition, paramagnetic behavior due to the Ni atoms has been observed in $NiGa_2O_4$ crystals at room temperature and above[25]. While it is not possible to truly model a paramagnetic system with our unit cell size, we model three unique antiferromagnetic (AFM) spin configurations for each structure to get a sense of the influence of spin on band gap prediction. Ferromagnetic configurations were considered but are higher in total energy than AFM and thus not studied for all structures. While proper investigation of cation and spin disorder requires supercells and/or random sampling methods, our results provide sufficient indication of the effects of disorder on the crystallographic and electronic structures. The calculated lattice constants for $NiGa_2O_4$ $P4_122$ converted into dimensions of the cubic $Fd3m$ for easier comparison to measured values of nanocrystals[25] are in good agreement with experiments, as shown in Table S2.

The electronic structure calculations for various structures of spinel $NiGa_2O_4$ show that its direct band gap is influenced by cation ordering. As shown in Fig. 2b, the lowest energy configuration is the ordered fully inverse spinel with *$P4_122$* crystal space group. The highest energy and largest bandgap configuration is the ordered normal spinel. The preference for inverse spinel as ground state (versus normal spinel) has been observed before[26] . As cation disorder is introduced, the band gap decreases, thus the set of $\lambda = 0.75$ configurations have the smallest band gaps. Even within the fully inverse spinel structures ($\lambda = 1$), this influence of disorder also appears, with GGA+U band gaps spanning 0.64 eV. Considering the various AFM spin configurations, bandgaps vary by about 0.23 eV. Combined, this could account for a significant range of possible measured bandgaps, depending on the degree of inversion and disorder in our samples.

To get a more accurate measure of the range of band gaps expected, we performed the HSE06+$G_0W_0$ calculations on three example structures: the largest band gap normal spinel with AFMv2, the lowest energy *$P4_122$* inverse spinel with AFMv1, and the smallest band gap inverse spinel ($\lambda = 1$) which we call inverse#10 with AFMv1. The HSE06+$G_0W_0$ band gaps are 4.64 eV for normal spinel, 4.44 eV for *$P4_122$* inverse spinel, and 3.73 eV for inverse#10. The range of predicted band gaps (Fig. 2b) overlap with the upper range of measured optical band gaps from Fig. 2a. These results suggest that our PLD growth techniques introduce disorder for the deposited $NiGa_2O_4$ films, especially for lower substrate temperatures.

**Thermal conductivity and phonon scattering of $NiGa_2O_4$ on $Al_2O_3$**

TDTR measurement results show that the cross-plane thermal conductivity of $NiGa_2O_4$ layers on $\alpha$–$Al_2O_3$ varies based on growth temperature. As shown in Fig. 3a, thermal conductivity of $NiGa_2O_4$ increases from 4 to 9 W $m^{-1}$ $K^{-1}$ as the substrate temperature setpoint is raised from 300 to 900 °C. A representative fit of the $NiGa_2O_4$ experimental data to the thermal model is shown in Fig. S7. This improvement in thermal conductivity may be attributed to enhanced cation ordering within the spinel lattice, aligning with the increased optical absorption onset observed over the same temperature range (Fig. 2a). This result is consistent with prior simulations of the phonon thermal conductivity of chemically ordered lattices[27–29], and the experimental measurements of the thermal conductivity of metals with varying degrees of chemical order[30,31].

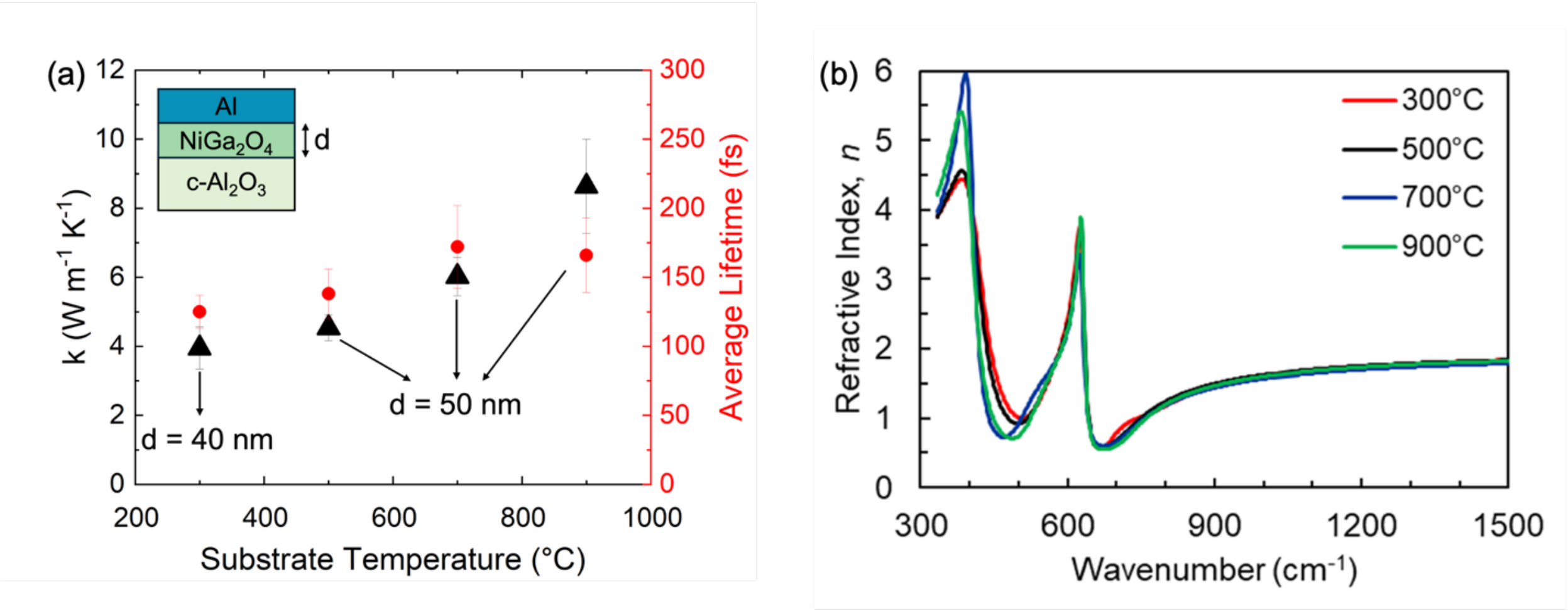

Fig. 3. (a) Thermal conductivity and phonon lifetime of $NiGa_2O_4$(111)/ α-$Al_2O_3$(00.1) sample shown in the inset at different substrate temperature setpoints, and (b) refractive index determined from IR-VASE for samples grown at different temperatures, with sharp phonon peaks at low wavenumber. An increase in thermal conductivity (black solid triangle) and the average lifetime of phonons (red solid circle) is observed with increasing growth temperature. The error bars in (a) indicate the spot-to-spot variation in thermal conductivity and the uncertainty in phonon lifetime measurements.

The phonon lifetimes extracted from IR-VASE measurements (Fig. 3b) provide further insight into the lattice dynamics and fundamental mechanisms driving the thermal conductivity trends in $NiGa_2O_4$ thin films grown at different α–$Al_2O_3$ substrate temperatures. Fig. 3b shows the refractive index of $NiGa_2O_4$ in the IR range, where sharp peaks at low wavenumber correspond to the phonon vibrations (see Fig. S8 and Fig. S9 for raw data and more detailed analysis). The width of these peaks decreases with increasing substrate temperature, indicating higher phonon lifetimes. As shown in Fig. 3a, the average lifetimes extracted from IR-VASE (Fig.3b) increase with deposition temperature, indicating a reduction in phonon scattering as the crystalline order improved. The films deposited at lower temperatures exhibited shorter lifetimes, consistent with higher cation disorder and stronger phonon-defect interactions. The films grown at higher temperatures showed longer lifetimes, reflecting enhanced structural uniformity and reduced lattice imperfections.

When compared with the thermal conductivity values, the phonon lifetime data show a direct and proportional relationship (Fig.3a). Samples with longer phonon lifetimes exhibit higher cross-plane thermal conductivity as both quantities are governed by the same underlying phonon scattering mechanisms. The enhancement of both lifetime and thermal conductivity with increasing deposition temperature highlights that both may be related to the improved cation ordering within the spinel lattice. These findings confirm that optimizing growth conditions to reduce disorder not only improves the electronic and optical quality of $NiGa_2O_4$ but also directly strengthens its phonon-mediated thermal transport, a key property for high-power and high-temperature oxide electronics.

**Epitaxial growth and devices of spinel $NiGa_2O_4$ on β-$Ga_2O_3$ substrates**

To determine the $NiGa_2O_4$ - $Ga_2O_3$ epitaxial alignments, $NiGa_2O_4$ was deposited at 600°C on $Ga_2O_3$ substrates with (100) and (010) orientations. Ultrathin $NiGa_2O_4$ layers grown on $Ga_2O_3$ (100) substrates exhibit coherent epitaxial strain with critical thicknesses of up to 10 nm, making them suitable as interfacial layers in $Ga_2O_3$/NiO junctions. The strain and the relaxation was measured by reciprocal space mapping using the (910) $Ga_2O_3$ and (622) $NiGa_2O_4$ reflections in Fig S3. As shown in Fig. 4a, these $NiGa_2O_4$ layers have coincident lattice alignment with the $Ga_2O_3$ substrate at a 45° rotation of the spinel lattice giving an in-plane orientation $Ga_2O_3$ [010] || $NiGa_2O_4$ [011] (Fig. S3). This lattice alignment corresponds to mismatch of 1.6% for the $NiGa_2O_4$ (100) grown on $Ga_2O_3$(100) substrates. XRD data demonstrate the (400) reflection peak of the thinnest 3-5 nm $NiGa_2O_4$ layers to be coherent with the substrate lattice, appearing above 44° degrees.

As layer thickness increases to 8–10 nm, the peak intensity grows, the peak positions shift to lower angle, and two distinct peaks components emerge, suggesting partial relaxation of the $NiGa_2O_4$ film. The wide-angle XRD scan for this sample and the rocking curve measurement with the FWHM values of 0.48 for the (400) reflection are shown in Fig. S2. In comparison, the $NiGa_2O_4$ (220) thin film on $\beta$-$Ga_2O_3$ (010) substrate has wider rocking curve with the FWHM values of 0.82 for the (440) reflection in the XRD (Fig. S4). The high-resolution scan of the $NiGa_2O_4$ (220) thin film on β-$Ga_2O_3$ (010) substrate is presented in Fig. S5.

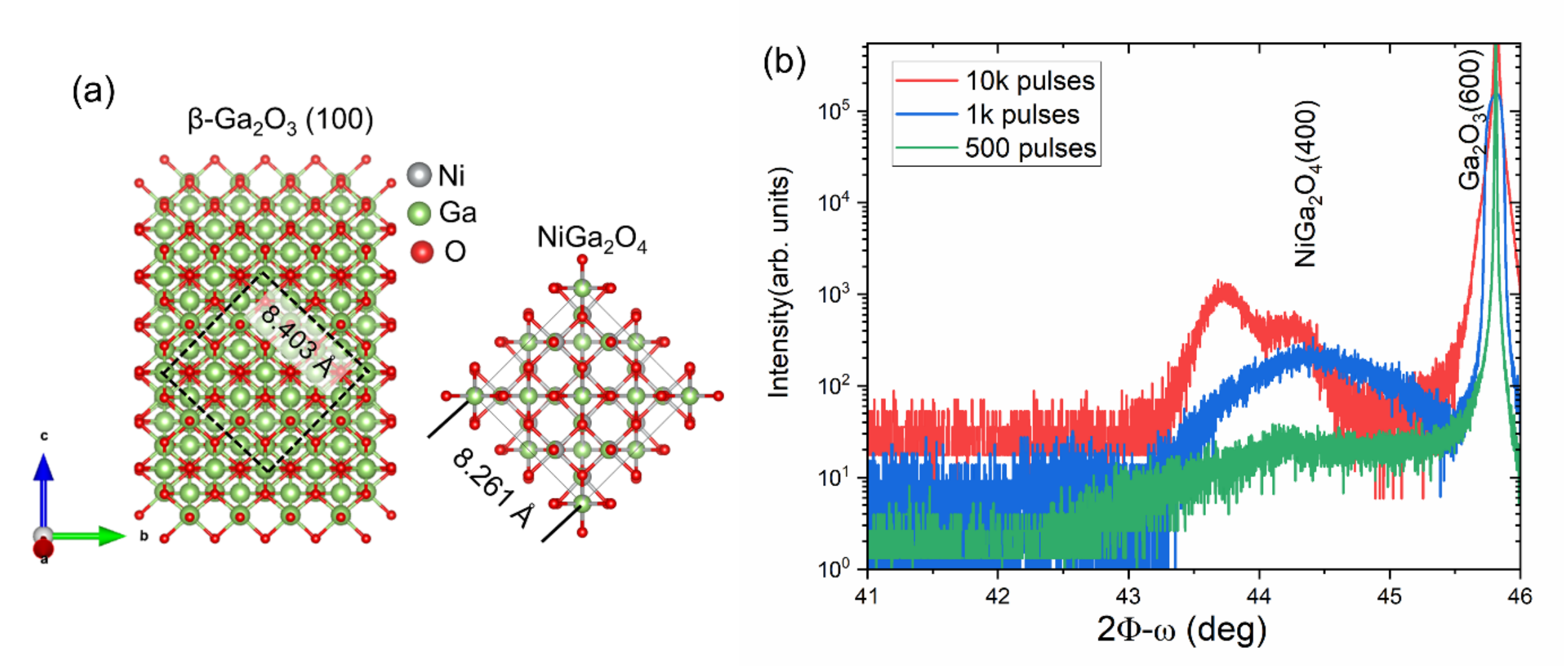

Fig. 4. (a) Monoclinic β-$Ga_2O_3$ projected along the normal to (100) showing a cubic co-incident lattice with lattice constant of 8.403 Å and $NiGa_2O_4$ unit cell in the (100) direction rotated by 45 degrees with lattice constant, a=8.261 Å. (b) Out-of-plane 2θ-ω scan showing heteroepitaxial spinel-type $NiGa_2O_4$(400) on β-$Ga_2O_3$(100) for films grown at 700 °C for different number of pulses corresponding to different thicknesses in the 3-10 nm range.

Fig. 5a shows the measured room temperature I-V characteristics of the $NiGa_2O_4/Ga_2O_3$ (001) heterojunction diode for different contact diameters. The inset shows the schematic of the device stack on the $Ga_2O_3$ substrate with a 1 μm thick epitaxial layer commercially grown by HVPE. An excellent rectification ratio ~ $10^9$ is obtained for this diode device at ambient temperature. The temperature dependence of these J-V curves for the 300 μm diameter pad is shown in Fig.5b. The rectification ratio of 10-100 is maintained even at the highest 600 °C measurement setpoint temperature, indicating good thermal reliability of the interface.

Reverse bias leakage current mechanisms were analyzed for $NiGa_2O_4/Ga_2O_3$ heterojunction device using the thermionic emission and Poole-Frenkel models, as detailed in the Methods section. The device measurement and modeling parameters for the $NiGa_2O_4/Ga_2O_3$ p-n heterojunction devices are summarized in Table S3. The ideality factor (η=1.6), turn on-voltage (V=1.5V), barrier height ($\Phi_B$ = 1.6 eV), and the on-state resistance ($R_{on}$ = 49 mΩ-$cm^2$) are extracted from the device model. The thermionic emission model shows a lower barrier height (1.12 eV) and has a better fit compared to PF model (1.32 eV), suggesting that for higher temperatures thermionic emission is the dominant mechanism.

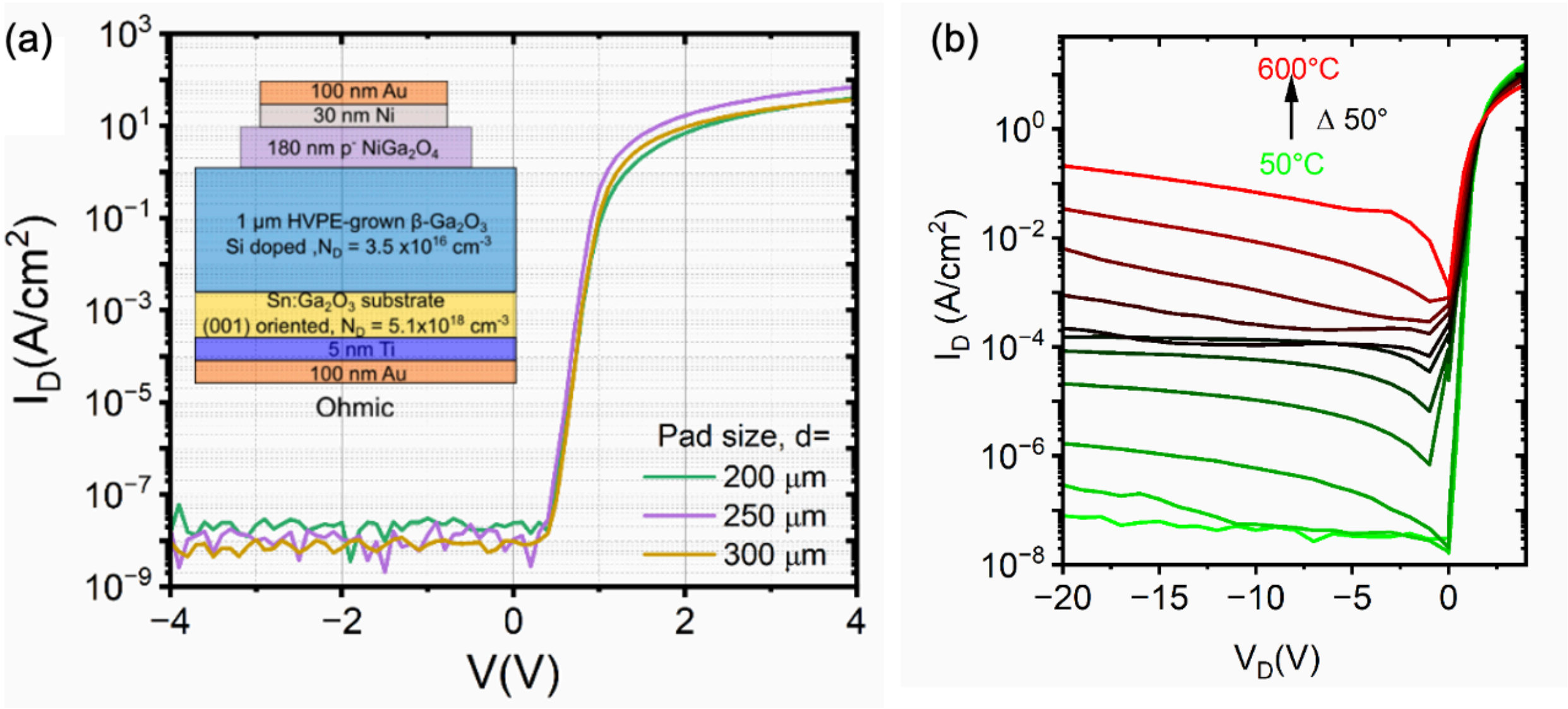

Fig. 5: (a) Room temperature I-V characteristics of the $NiGa_2O_4/Ga_2O_3$ device for different pad sizes. Inset shows a schematic of the device structure. (b) Temperature dependent I-V characteristics (I-V-T) of a 300 μm diameter *p*-$NiGa_2O_4$/*n*-$Ga_2O_3$ heterojunction diode measured between 50 to 600 ºC.

**$NiGa_2O_4/Ga_2O_3$ band offset measurements**

The Kraut method, that is routinely used to characterize VBOs at semiconductor interfaces, cannot be straightforwardly applied to interfaces like $NiGa_2O_4$ / $Ga_2O_3$, where no unique substrate core level exists. The Kraut method requires a minimum of three samples: a substrate sample; a thin (~3-5 nm, less than XPS information depth) overlayer deposited on the substrate; and a thick (>10 nm, greater than XPS information depth) overlayer sample. The substrate and thick overlayer samples provide valence-band maximum (VBM) values and core-level positions for unique elemental constituents in the substrate and overlayer. From these, core-level-to-VBM energy separations characteristic of the pure materials are extracted. The thin deposited sample provides substrate-to-overlayer core-level energy separations, which when combined with the core-level-to-VBM values, allows extraction of the interfacial VBO.

To address the challenge of the $NiGa_2O_4$ / $Ga_2O_3$ band offset measurements, we developed and applied a new method for VBO determination based on curve fitting both valence-band and core-level XPS spectra. We acquired XPS spectra on sample series for both $Ga_2O_3$ (100) and $Ga_2O_3$ (001) substrates: the bare substrates, and in this case three successively thicker $NiGa_2O_4$ / $Ga_2O_3$ samples. To extract VBOs, valence-band spectra for $Ga_2O_3$ substrates and thickest $NiGa_2O_4$ samples were first fitted with gaussian components. The fits were subsequently interpolated onto a finer binding-energy scale (0.01 eV vs. 0.05 eV for the native spectra). The curve fitting removes noise from the spectra without spectral-feature broadening that occurs with smoothing algorithms, while the interpolation facilitates smaller energy shifts in the fitting process used to extract VBO values. Best fits to valence-band spectra for $Ga_2O_3$ substrates and thickest $NiGa_2O_4$ overlayer films are shown in Fig. S10.

The extract the VBO from valence-band spectra, a linear combination of the interpolated, fitted valence-band spectra for the $Ga_2O_3$ substrates and thickest $NiGa_2O_4$ samples were used to fit the spectra from the thin $NiGa_2O_4$ / $Ga_2O_3$ samples. Adjustable fitting parameters include amplitude factors and binding-energy shifts relative to as-acquired spectra from the $Ga_2O_3$ substrates and thickest $NiGa_2O_4$ film samples. An additional fitting parameter set the fraction of the valence-band spectrum for the thickest $NiGa_2O_4$ film that could be attributed to signal from the $Ga_2O_3$ substrate, as shown by analysis and fitting of the core level spectra (Fig. S11). This fitting parameter was set to 0.50 based on analysis of the Ga $2p_{3/2}$ and O 1s peak core-level peak-area ratios. Relationships between measured and extracted valence-band spectra are:

$$VB_{NiGa_2O_4} = VB^{sh}_{thick\ NiGa_2O_4} - 0.5 \cdot VB^{sh}_{Ga_2O_3} \quad (4)$$

$$VB_{thin\ NiGa_2O_4} = a \cdot VB^{sh}_{Ga_2O_3} + b \cdot VB_{NiGa_2O_4} \quad (5)$$

where superscript “sh” refers to binding-energy shifts in each spectrum that arise from substrate band bending.

Fig. 6a presents XPS measurement results, where the deconvoluted spectra show separation of $Ga_2O_3$ and $NiGa_2O_4$ contributions (Fig. S10). Using this new analysis approach, the valence band (VB) offset between $NiGa_2O_4$ and $Ga_2O_3$ is estimated to be 1.8 eV, with an additional 0.4 eV due to band bending observed as the $NiGa_2O_4$ overlayer grows for the $Ga_2O_3$ (100) orientation. The measured XPS valence band offset for the thin $NiGa_2O_4$/ β-$Ga_2O_3$ heterostructure with (001) orientation is similar, but without the band bending, as shown in Fig. S12.

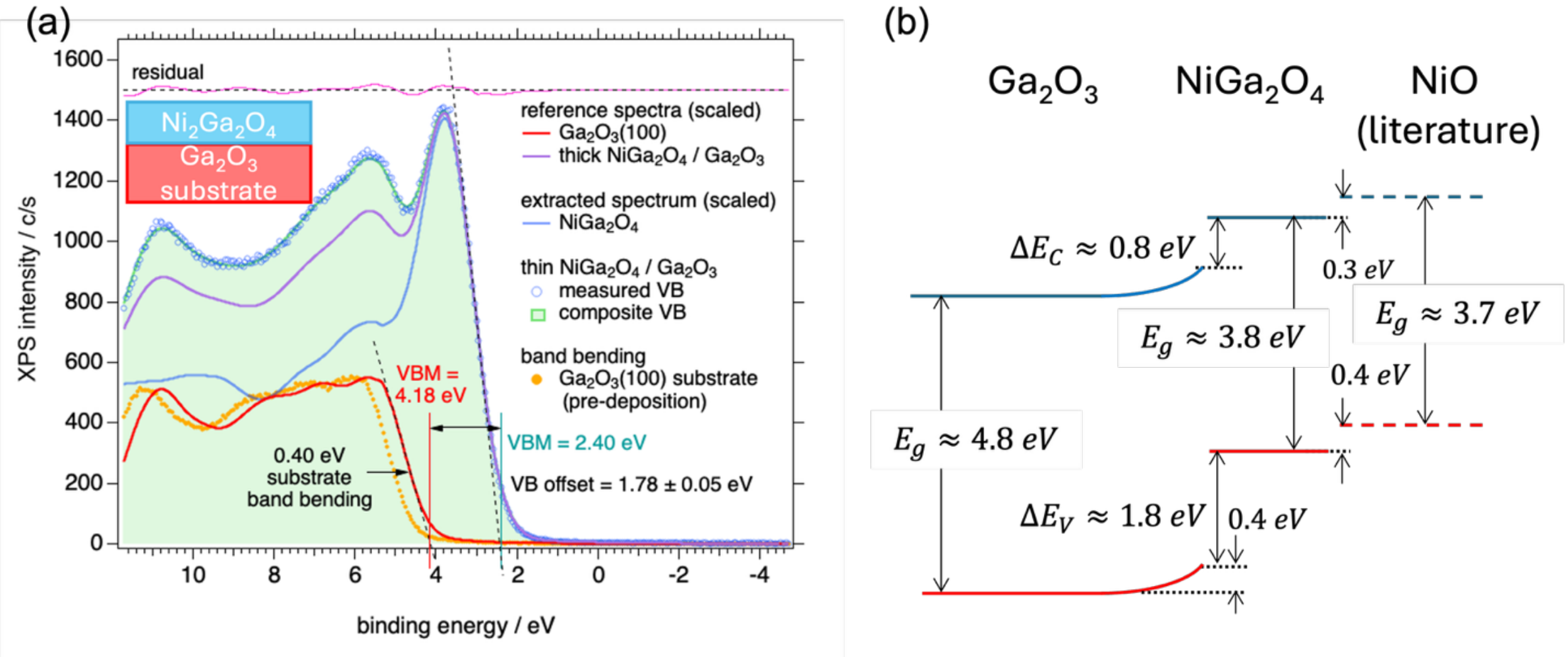

Fig. 6: (a) XPS valence band spectra of thin $NiGa_2O_4$/ β-$Ga_2O_3$(100) heterostructure showing extrapolated VB offset of ~ 1.8 eV and 0.4 eV band bending in the $Ga_2O_3$ substrate. (b) Schematic band diagram (assuming flat band) of the NiO/$NiGa_2O_4$/$Ga_2O_3$ heterostructure using NiO experimental data obtained from literature[32] .

These band offset measurement results are summarized by a band diagram in Fig. 6b, taking into account the measured bandgap of 3.8 eV for $NiGa_2O_4$ and the established band gap of 4.8 eV for $Ga_2O_3$. For the measured valence band offset of 1.8 eV (not taking 0.4 eV band bending into account), the conduction band offset between these two materials is determined to be 0.8 eV. The valence band offset for the $NiGa_2O_4$ / NiO interface is derived to be 0.4 eV, taking into account the literature data for the $Ga_2O_3$/NiO interfaces[32] .

It is notable that this fitting approach allows the valence-band spectrum characteristic of $NiGa_2O_4$ and the $NiGa_2O_4$ / $Ga_2O_3$ to be determined despite the $Ga_2O_3$ substrate had no unique core level peaks, and even though $NiGa_2O_4$ samples thicker than the XPS information depth were not available in this study. This $NiGa_2O_4$ / $Ga_2O_3$ example illustrates the advantage of this new method to analysis of chemically convoluted interfaces used in power electronics and other applications.

## SUMMARY AND CONCLUSION

In this paper, structural, optical and thermal properties of $NiGa_2O_4$ thin films grown are investigated as p-type interlayers forming a rectifying $Ga_2O_3$/NiO junction for vertical device applications. Pulsed laser deposition enabled epitaxial growth of high-quality cubic spinel $NiGa_2O_4$ films on $\alpha$-$Al_2O_3$(00.1) and monoclinic $\beta$-$Ga_2O_3$ (100), (010) and (001) substrates. Epitaxial $NiGa_2O_4$ layers are shown to have low surface roughness (~2 nm), optical absorption onset of 3.6-3.9 eV, and thermal conductivity of 4-9 W/mK, as the substrate temperature increases from 300 to 900 °C during the growth. Infrared ellipsometry further indicated increasing phonon lifetimes with increasing deposition temperature, correlating with the increase in thermal conductivity. First-principles calculations show that variations in the Ni and Ga cation ordering in the spinel lattice can account for these temperature-dependent optical and thermal trends. The valence band offset between $NiGa_2O_4$ and $\beta$-$Ga_2O_3$ was determined to be 1.8 eV, with an additional band bending of 0.4 eV in $Ga_2O_3$ (100) substrate, measured via X-ray photoelectron spectroscopy and valence band fitting analysis. The $NiGa_2O_4$/$Ga_2O_3$ (001) p-n heterojunction device exhibited a rectification ratio of $10^8$ (at $\pm$2V) and a turn-on voltage of 1.4 V, maintaining diode behavior up to 600 ºC. These results suggest that $NiGa_2O_4$ spinel layers at $Ga_2O_3$/NiO interfaces can provide a chemically compatible, thermally stable, and electronically favorable interfacial solution for advanced electronic devices operating at high-power and high-temperature conditions. Together, this paper shows how the unintentional interfacial layers that form at the heterojunction of devices operating in extreme conditions can be studied in isolation to understand and control their properties for the device performance improvements.

## METHODS

### Materials growth

$NiGa_2O_4$ thin films were grown on (100) and (010) Sn-doped $\beta$-$Ga_2O_3$ substrates (Novel Crystal Technology) and $\alpha$-$Al_2O_3$(00.1) substrates by PLD. Commercial $NiGa_2O_4$ ceramic targets (Plasmaterials, 99.99%) were ablated at a frequency of 10 Hz using a KrF excimer laser (248 nm). Before growth the $\beta$-$Ga_2O_3$ substrates were cleaned in a Piranha solution followed by acetone and IPA rinse and pretreated by annealing at 800 °C for 1 hour in a box furnace in laboratory ambient. Before growth, the chamber was evacuated to a based pressure of ~ $10^{-8}$ mTorr and the substrate was heated to the growth temperature in vacuum at a rate of 20°C /min, and then high purity $O_2$ gas was used to achieve a growth pressure of ~$1x10^{-3}$ mTorr. The growth rate and film thickness were controlled by the number of pulses. The films were cooled postgrowth at the rate of 10°C/min at the growth $O_2$ partial pressure. In-situ RHEED patterns during and after growth were obtained using a double differential pumped RHEED system (TorrRHEED™, Staib Instruments).

**Materials characterization**

Wide angle out-of-plane diffraction, phi-scans and the reciprocal space maps were obtained by high resolution X-ray diffraction (XRD) using a Rigaku SmartLab instrument equipped with a Ge (220) x2 double bounce monochromator with a Cu Kα radiation source. A Bruker D8 Advance diffractometer was also used to obtain 2-dimensional (2θ, χ) XRD frame of the thin films. Optical properties and bandgaps of the thin films were measured by standard (isotropic) Spectroscopic ellipsometry (SE). SE measurements were taken in the spectral range of 0.73 eV to 6.5 eV using a rotating-compensator instrument (J. A. Woollam, M-2000). The angle of incidence varied from 55° to 75° with an increment of 10°. In the SE analysis to obtain the optical constants of $NiGa_2O_4$, we assumed an optical model comprising a surface roughness/$NiGa_2O_4$ thin films/$\beta$-$Ga_2O_3$ substrate structure. The Bruggeman effective medium approximation (EMA) with a 50:50 vol. % mixture of the bulk layer, and void was used to model the optical properties of the surface roughness layer.

**Thermal conductivity measurements**

Thermal properties of 40 nm and 50 nm thick $NiGa_2O_4$ on $\alpha$-$Al_2O_3$(00.1) substrate with a growth temperature of 300, 500, 700 and 900 °C were measured using TDTR in a two-tint configuration[33,34]. In this setup, a sub-picosecond laser pulse output from a Ti-sapphire oscillator (80 MHz repetition rate) is split into a pump and probe path using a polarized beam splitter (PBS). The pump is modulated with an electro-optic modulator (EOM) at frequencies ranging from 1.1 MHz to 8.4 MHz. The probe pulses are delayed with respect to that of the pump with a mechanical delay line. The pump and probe are focused on the sample using a 10X objective resulting in $1/e^2$ beam spot radii of ~ 6 $\mu$m. The pump induces a localized heating event, which in turn causes a change in reflectance at the sample surface and therefore the intensity of the reflected probe beam changes at modulation frequency of the pump heating event. To ensure linearity in the thermoreflectance signal, dR/dT, and that the change in reflectivity is related to the surface temperature change, an Al transducer is applied[33] . The reflected probe beam is monitored using a lock-in detection scheme with $V_{in}$ (in-phase) and $V_{out}$ (out-of-phase) signals as outputs[33]. The negative ratio of $V_{in}$ and $V_{out}$ is fit with a heat diffusion model to extract the unknown thermal properties[35] . In a TDTR measurement, the fit parameters are pre-determined through sensitivity analysis[33,34] . In this study, based on the pump modulation frequency of 8.4 MHz and 1.1 MHz, the $-V_{in}/V_{out}$ signal sensitivity is sufficient to fit for the cross-plane thermal conductivity of $NiGa_2O_4$, the $\alpha$-$Al_2O_3$ substrate, and the thermal boundary conductance (TBC) at the interface of Al/$NiGa_2O_4$. The heat capacities for the Al transducer, $NiGa_2O_4$ and the $\alpha$-$Al_2O_3$ are obtained from literature[33,36]. The thermal conductivity and thickness of the Al transducer is obtained from four-point probe and picosecond ultrasonics, respectively[37,38].

**Phonon scattering measurements**

Infrared variable-angle spectroscopic ellipsometry (IR-VASE, J.A. Woollam IR-VASE Mark II) was used to characterize the phonon response and dielectric function of $NiGa_2O_4$ thin films grown at different deposition substrate temperatures. Measurements were conducted at room temperature over the spectral range of 333–5000 $cm^{-1}$ (2-30 μm) using incidence angles of 60° and 70°. The ellipsometric parameters $\Psi$ and $\Delta$, were analyzed using a multilayer isotropic optical model consisting of a surface roughness layer, the $NiGa_2O_4$ film, and the underlying $\alpha$-$Al_2O_3$ substrate. Surface roughness was described using the Bruggeman effective medium approximation with a 50:50 vol.% mixture of film and void. The complex dielectric function, $\varepsilon(\omega) = \varepsilon_1(\omega) + i\varepsilon_2(\omega)$, was modeled as a sum of Lorentz oscillators representing infrared-active phonon modes:

$$\varepsilon(\omega) = \varepsilon_\infty + \sum_j \frac{f_j}{\omega_j^2 - \omega^2 - i\Gamma_j\omega} \quad (1)$$

where $f_i$, $\omega_i$, and $\Gamma_i$ are the oscillator strength, resonance frequency, and damping coefficient of the $j^{th}$ mode, respectively. Model parameters were optimized by minimizing the mean-square error between measured and calculated $\Psi$ and $\Delta$ spectra. The phonon lifetime ($\tau_j$) associated with each oscillator was determined from its linewidth according to $\tau_j = (2\pi c\Gamma_j)^{-1}$, where c is the speed of light. Weighted averages of $\tau_j$, based on oscillator strengths, were calculated for each film to represent the effective optical phonon lifetime, which reflects the rate of optical phonon scattering and lattice disorder introduced by growth conditions.

**Photoelectron spectroscopy measurements**

Core-level and valence-band XPS measurements were performed to characterize $NiGa_2O_4$ / $Ga_2O_3$ valence-band offsets (VBOs), chemical states, and compositions. Measurements were performed with a Physical Electronics PHI 5000 VersaProbe III instrument using monochromatic Al-kα (hν = 1486.6 eV), 55.0 eV pass energy, and a 5° exit angle with respect to the surface normal. The XPS core-level spectra from $Ga_2O_3$ substrates and epitaxial $NiGa_2O_4$ / $Ga_2O_3$ samples were fitted using the methodology published earlier[39], and analyzed using the new method described in the results section of publication that works for materials with common elements unlike the conventional Kraut method[40]. In this approach, fitted spectra for the bare $Ga_2O_3$ substrates provide constraints on Ga $2p_{3/2}$ -to- O 1s binding-energy separations and peak-area ratios. Similarly, tabulated elemental sensitivity factors for Ni $2p_{3/2}$, Ga $2p_{3/2}$, and O 1s constrain relative peak-area ratios for the $NiGa_2O_4$ phase. The results of the XPS core level analysis of $NiGa_2O_4$ on $Ga_2O_3$ substrates indicated the presence of $Ni(OH)_2$ and C-containing species at the surface.

**Device fabrication**

For the fabrication of $NiGa_2O_4/Ga_2O_3$ (001) vertical heterojunction devices, a Ti/Au (5 nm/100 nm) ohmic metal stack was deposited by electron beam evaporation to cover the backside of the Sn-$Ga_2O_3$ wafer followed by a rapid thermal annealing (RTA) in $N_2$ ambient at 550 °C for 1 min. Contact aligner lithography was used to make circular contact patterns (50-300 $\mu$m diameter) on the top $NiGa_2O_4$ layer followed by e-beam evaporation of Ni/Au (30 nm/100nm). To extract the device properties, the fabricated diodes were characterized with current-voltage (J-V) measurements at room temperature using a Keithley 4200A semiconductor parameter analyzer. Temperature dependent I-V (I-V-T) measurements were conducted at ~$10^{-4}$ Torr using a custom McAllister vacuum probe station.

**Device analysis**

Reverse bias leakage current mechanisms were analyzed for the $NiGa_2O_4/Ga_2O_3$ heterojunction device using the thermionic emission and Poole-Frenkel models. Thermionic emission (TE) describes the thermal excitation of charge carriers over an electronic barrier:

$$J_0 \propto T^2 \exp\left(\frac{-\phi_{TE}}{k_B T}\right) \quad (2)$$

where $T$ is the absolute temperature, $\phi_{TE}$ is the effective barrier height, and $k_B$ is Boltzmann's constant (in eV/K). The barrier height for TE can be extracted from the slope of $\ln\left(\frac{J_0}{T^2}\right)$ $vs. \frac{1}{k_B T}$.

Poole-Frenkel emission (PFE) describes the conduction across an insulating material via thermal/field excitation from trap states. This mechanism has stronger dependence on electric field compares to the previous mechanism, and is typically observed in highly resistive semiconducting materials or in instances of high interfacial trap density.

$$J_0 \propto E \exp\left(\frac{-q(\phi_{PF} - \sqrt{qE/\pi\varepsilon_r\varepsilon_0})}{k_B T}\right) \quad (3)$$

To extract the barrier height, $\ln\left(\frac{J_0}{E}\right)$ vs. $\sqrt{E}$ is plotted, and the corresponding intercepts are plotted against $\frac{1}{k_B T}$, where the slope is the barrier height $\phi_{PF}$.

**Theoretical calculations**

To create the $NiGa_2O_4$ structures, we began with the 28-atom inverse spinel *Imma* crystal from the Materials Project[41] and swapped cation atom positions to enumerate all possible unique permutations of spinel. These structures sample degrees of inversion of λ = 0, 0.25, 0.5, 0.75, and 1. Degree of inversion λ is a dimensionless quantity, where the chemical formula is then written as $(Ga_\lambda Ni_{1-\lambda})[Ga_{2-\lambda}Ni_\lambda]O_4$, with parentheses and brackets representing the $Td$ and $Oh$ sites, respectively. For each structure we then modeled the three possible antiferromagnetic spin configurations (two spin up and two spin down nickels). Structures were relaxed with density functional theory using the GGA+U (generalized gradient approximation) exchange-correlation functional ($U$ = +3.0 for Ni)[42,43] as implemented in the Vienna ab-initio simulation package[44] . We used a 3x3x2 *k*-point mesh, a plane-wave cutoff energy of 340 eV, and a force convergence criterion of 5 meV/Å. Since GGA+U is known to underestimate electronic band gaps, we improve the fidelity of our electronic structure calculations for select configurations with an approach known as HSE06 + $G_0W_0$ [45–47]. In this approach, the hybrid functional, HSE06, is used to relax the structure and then one iterative step of the perturbative many-body *GW* approximation is performed using the HSE06 wavefunction. Using this method, we can improve the calculated lattice constant to within 1% of experiment (see Table S1) using the standard HSE06 mixing parameter (α=0.25) and obtain more accurate band gaps.

## ACKNOWLEDGEMENTS

This work was authored in part at the National Laboratory of the Rockies (NLR) for the U.S. Department of Energy (DOE) under Contract No. DEAC36-08GO28308. This work at NLR, UV, and CSM was primarily supported as part of APEX (A Center for Power Electronics Materials and Manufacturing Exploration), an Energy Frontier Research Center funded by the U.S. Department of Energy, Office of Science, Basic Energy Sciences under Award # ERW0345 (Materials synthesis and characterization, theoretical calculations). Device fabrication and band offset measurements were supported by the DOE's Office of Critical Minerals and Energy Innovation (CMEI), Advanced Materials & Manufacturing Technologies Office (AMMTO) program. Calculations were performed using computational resources sponsored by the DOE CMEI, located at the NLR. The authors also acknowledge Colorado School of Mines supercomputing resources (http://ciarc.mines.edu/hpc) made available for conducting the research reported in this paper. The views expressed in the article do not necessarily represent the views of the DOE or the U.S. Government.

## DATA AVAILABILITY

Data is available from corresponding author upon reasonable request.

## CONFLICT OF INTEREST

The authors declare no conflicts of interest

Supporting Information for

# Epitaxial growth and semiconductor properties of $NiGa_2O_4$ spinel for $Ga_2O_3$/NiO interfaces

Kingsley Egbo[1], Emily M. Garrity[2], Shivashree Shivamade Gowda[3], Saman Zare[3], Ethan A. Scott[3], Glenn Teeter[1], Brooks Tellekamp[1], Vladan Stevanovic[2], Patrick E. Hopkins[3], Andriy Zakutayev,[1†] Nancy Haegel[1]

[1] National Laboratory of the Rockies, Golden, CO 80401, USA

[2] Colorado School of Mines, Golden, CO 80401, USA

[3] University of Virginia, Charlottesville, VA 22904, USA

[†]Andriy.Zakutayev@nrel.gov

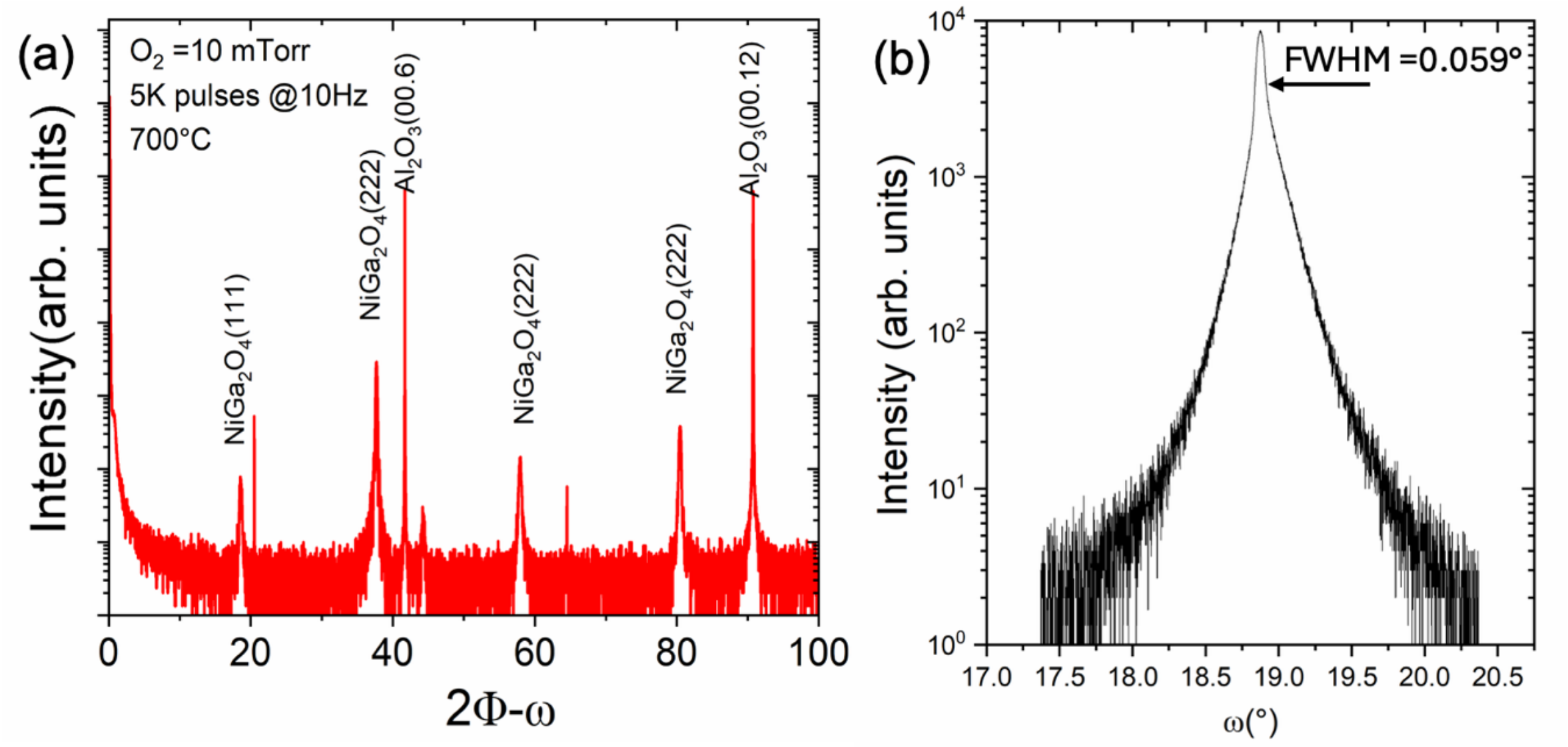

Figure S1: (a) Wide-angle XRD 2theta-omega scan for the $NiGa_2O_4$(111) thin film on α-$Al_2O_3$(00.2) deposited at $p(O_2)$ of 10 mTorr and substrate temperature of 700 °C. (b) Rocking curve measurement showing the FWHM values of 0.059 for the (222) reflection in the XRD 2theta-omega scan.

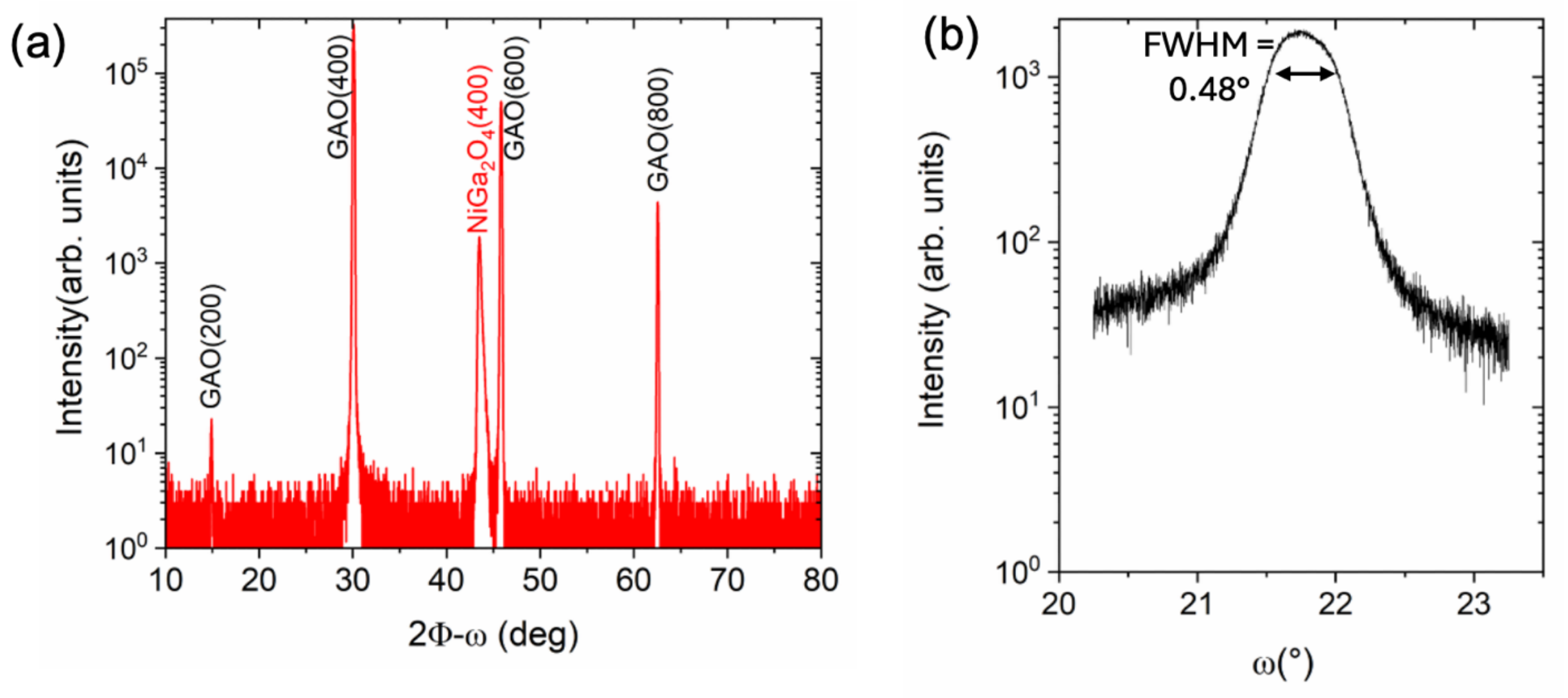

Figure S2: (a) Wide-angle XRD 2theta-omega scan for the $NiGa_2O_4$(200) thin film on *β*-$Ga_2O_3$(200) deposited at $p(O_2)$ of 10 mTorr and substrate temperature of 700 °C. (b) Rocking curve measurement showing the FWHM values of 0.48 for the (400) reflection in the XRD 2theta-omega scan.

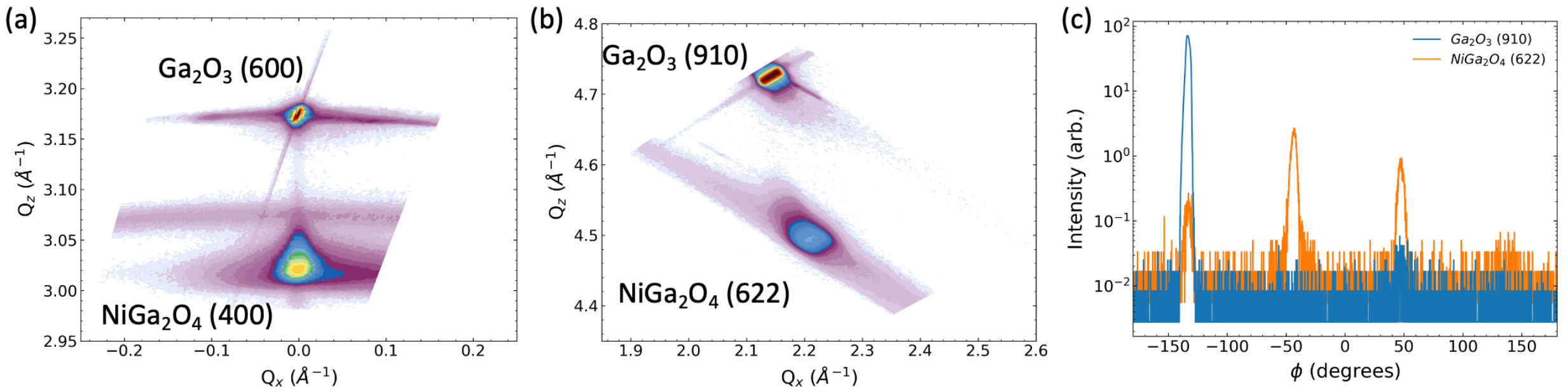

Figure S3: Reciprocal space maps (RSMs) of the 10k pulse (100) $NiGa_2O_4$ film grown on (100) $Ga_2O_3$. (a) symmetric RSM around the $Ga_2O_3$ (600) reflection and $NiGa_2O_4$ (400) reflection showing both strained and relaxed components of the epilayer. (b) asymmetric RSM around the the $Ga_2O_3$ (910) reflection and $NiGa_2O_4$ (622) reflection showing that the larger Qz component of the $NiGa_2O_4$ reflection, corresponding to the higher angle 2θ component, is strained to $Ga_2O_3$ while the larger intensity, smaller Qz component is partially relaxed and broadened along the omega tilt direction. (c) ϕ scan around the $Ga_2O_3$ (910) and $NiGa_2O_4$ (622) reflections demonstrating the in-plane epitaxial alignment (010) $Ga_2O_3$ || (011) $NiGa_2O_4$. The $Ga_2O_3$ peaks are similar, but not 4-fold symmetric resulting in nearly zero signal for all but the optimized peak.

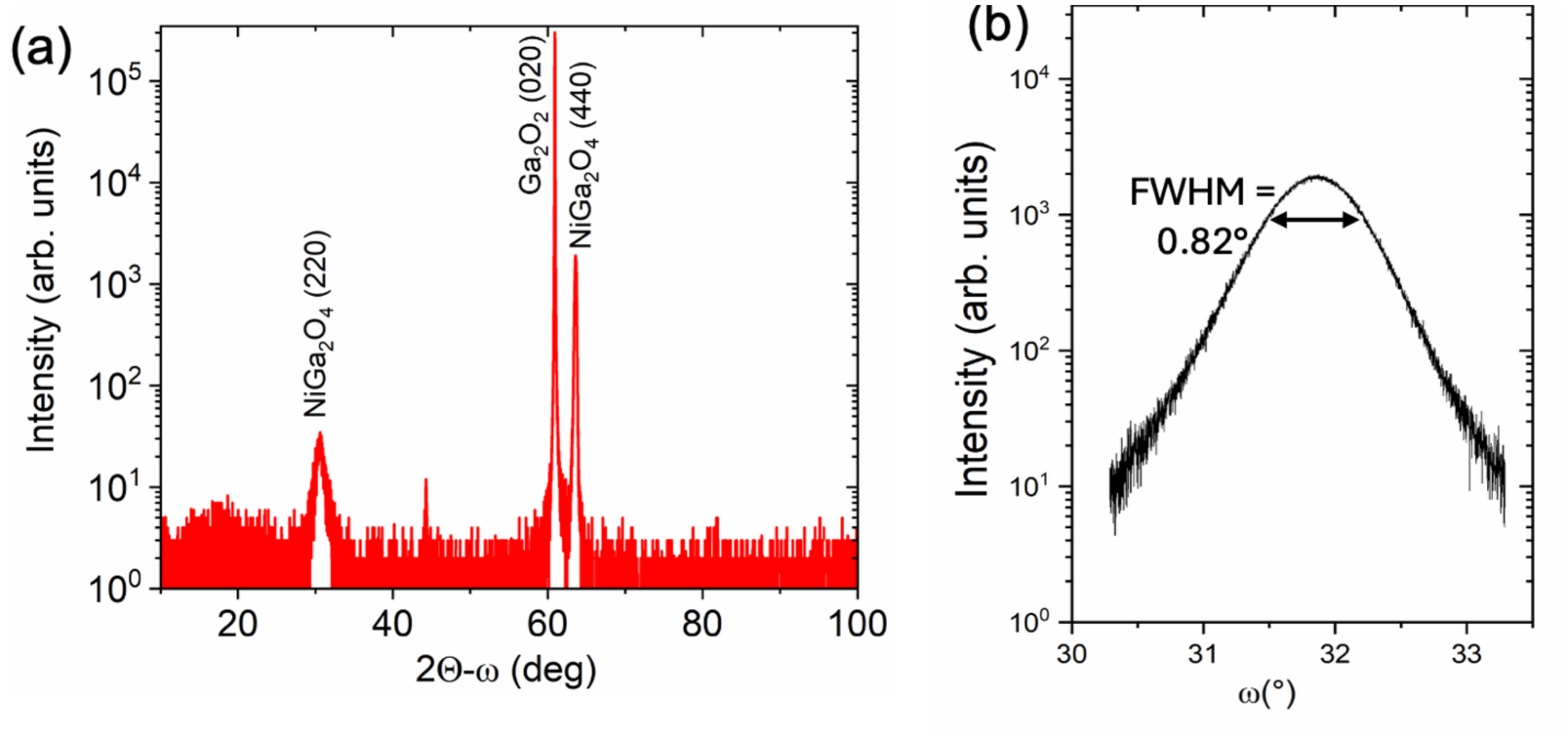

Figure S4: (a) Wide-angle XRD 2theta-omega scan for the $NiGa_2O_4$(220) thin film on *β*-$Ga_2O_3$(020) deposited at *p*($O_2$) of 10 mTorr and substrate temperature of 700 °C.(b) Rocking curve measurement showing the FWHM values of 0.82 for the (440) reflection in the XRD 2theta-omega scan.

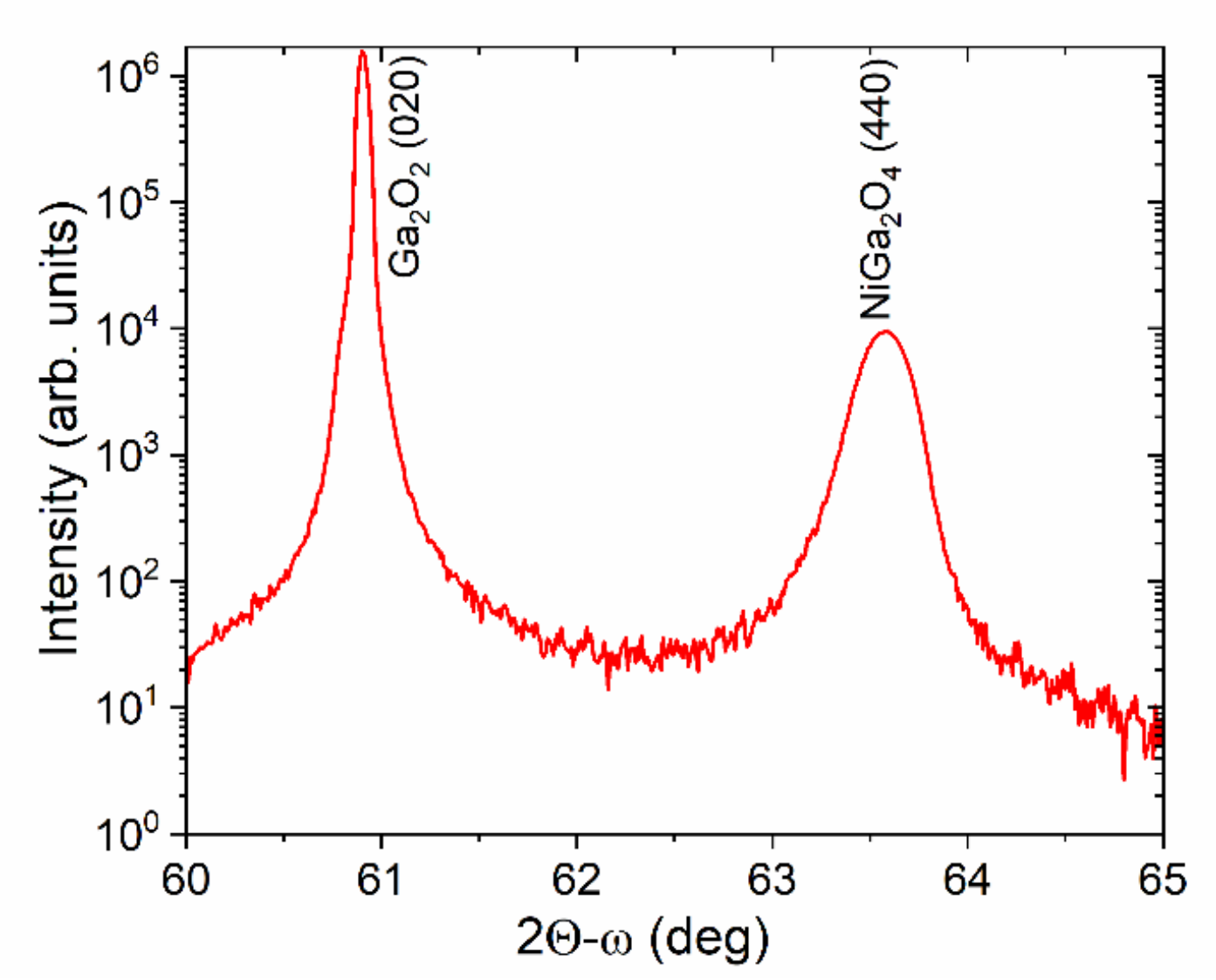

Figure S5: High resolution scan of the $NiGa_2O_4$(220) thin film on *β*-$Ga_2O_3$(020)

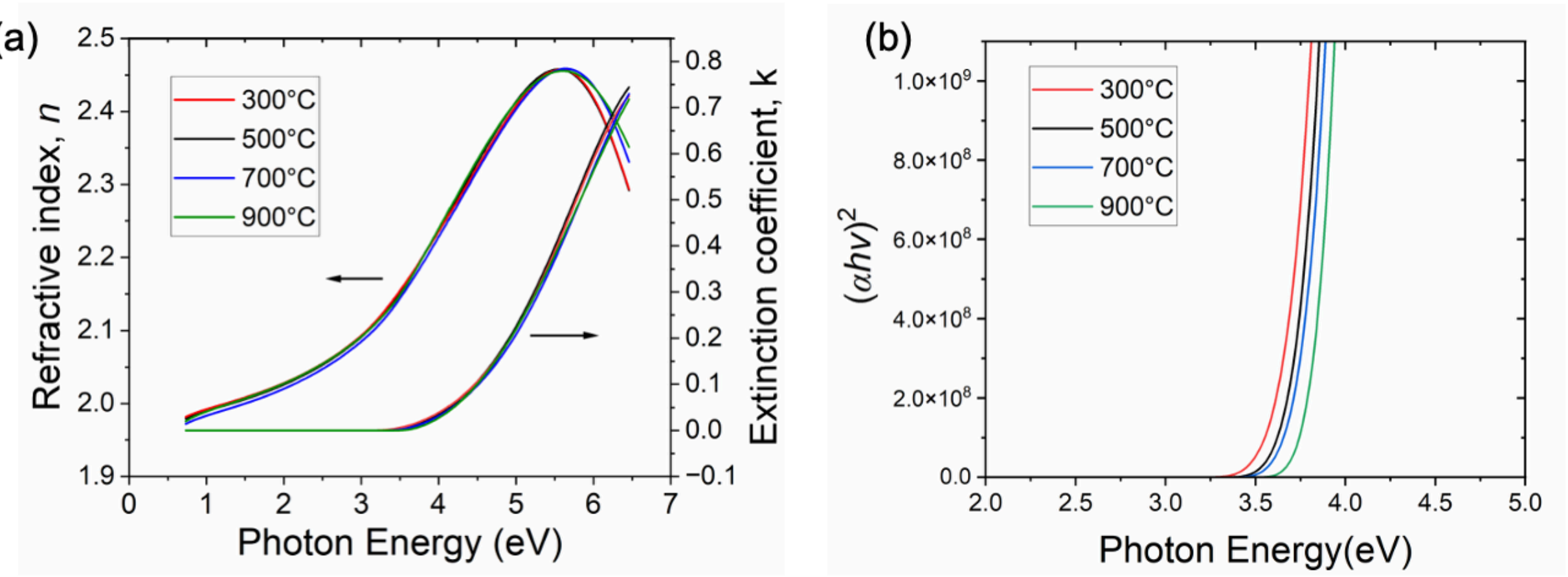

Figure S6: (a) Refractive Index (*n*) and extinction coefficient (k) spectra for the $NiGa_2O_4$(111) thin film on α-$Al_2O_3$(00.2) obtained from Spectroscopic Ellipsometry measurement and analysis in the spectral range of 0.7-6.5 eV. (b) Plot of $(\alpha hv)^2$ vs photon energy.

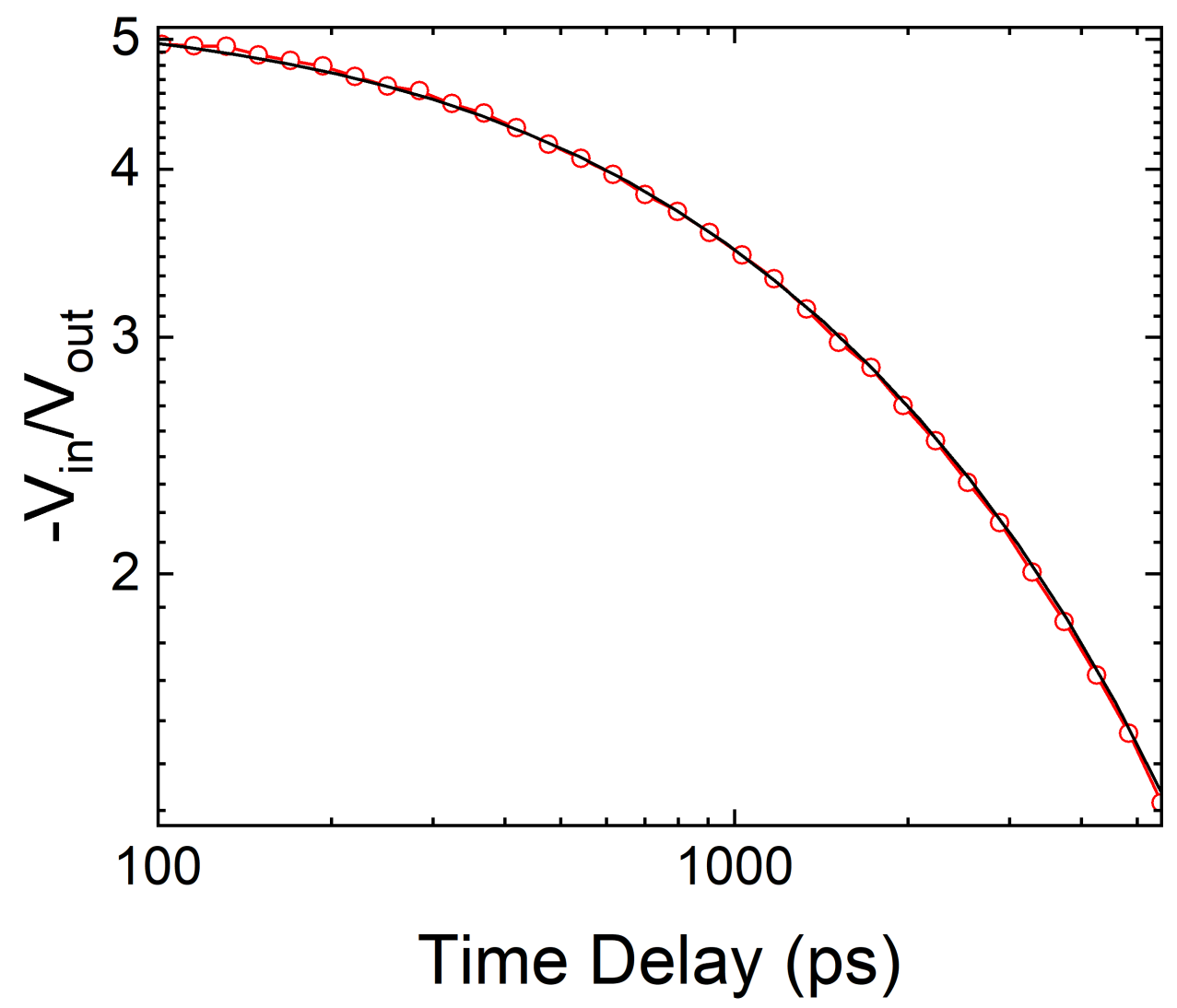

Fig. S7. TDTR $-V_{in}/V_{out}$ signal fitting of thermal model (solid black line) with the experimental data (open circles) across time delay for $Al/NiGa_2O_4/Al_2O_3$ at a modulation frequency of 8.4 MHz. In this fitting, the thickness of $NiGa_2O_4$ is 40 nm and the growth temperature is 300 ºC.

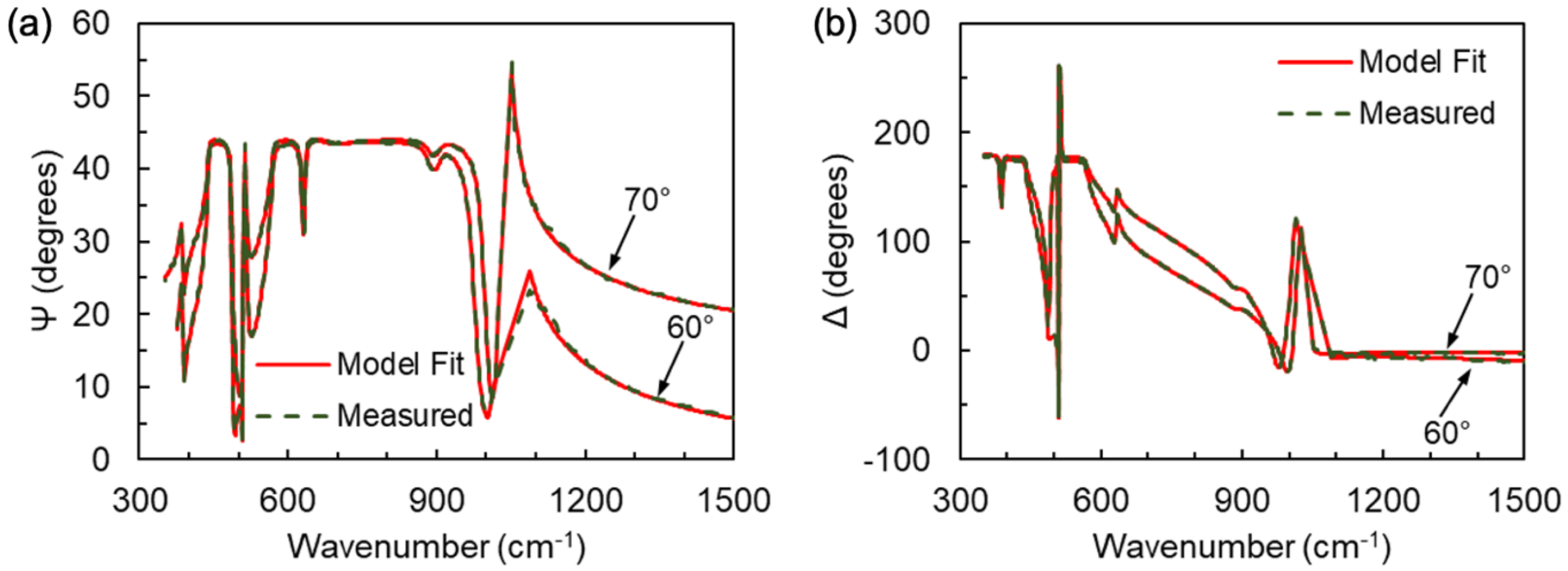

Figure S8: Measured IR-VASE data (Ψ and Δ) at the incidence angles of 60° and 70° for a substrate temperature of 300°C (green dashed lines), compared with modeled results (red solid lines) fitted to the measured data using the Lorentz model.

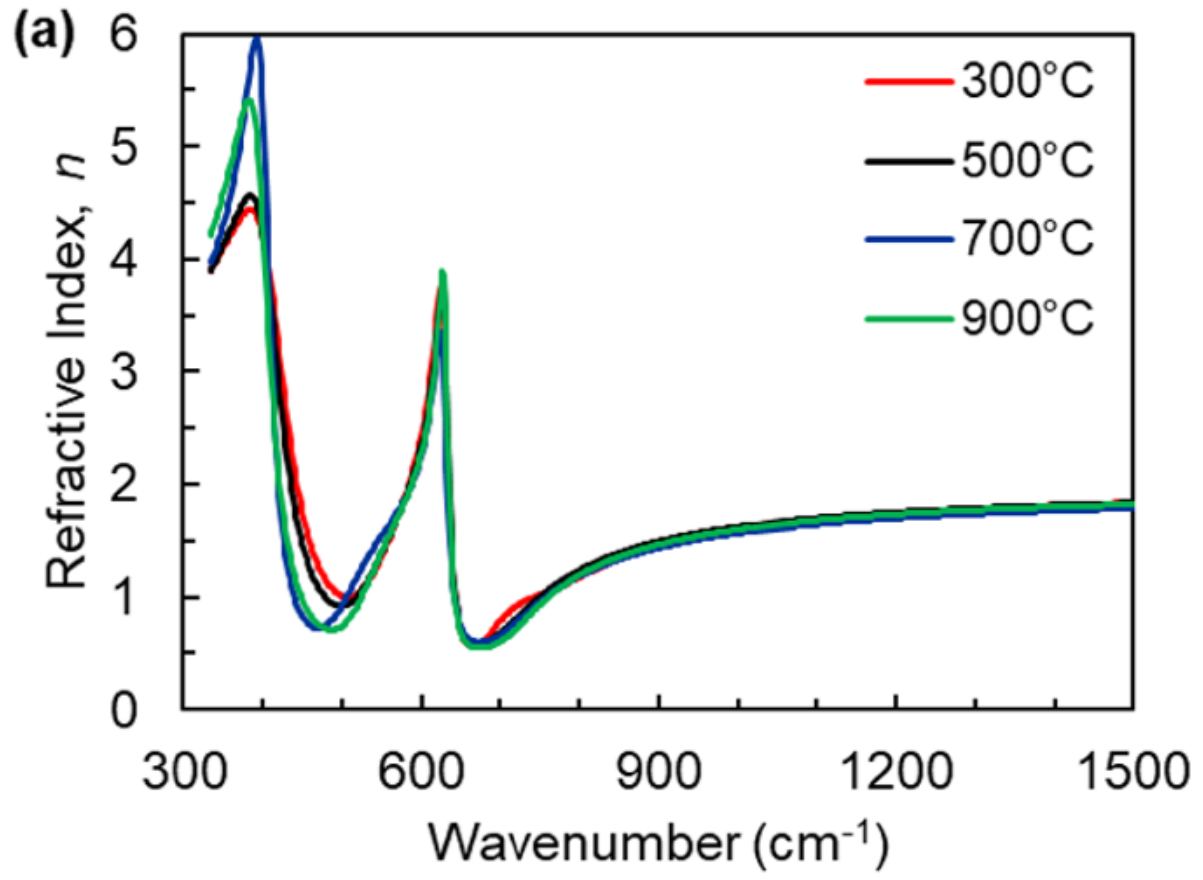

Figure S9: (a) Refractive index and (b) extinction coefficient spectra for the $NiGa_2O_4$ thin films at various substrate temperatures obtained from IR-VASE measurements and analysis.

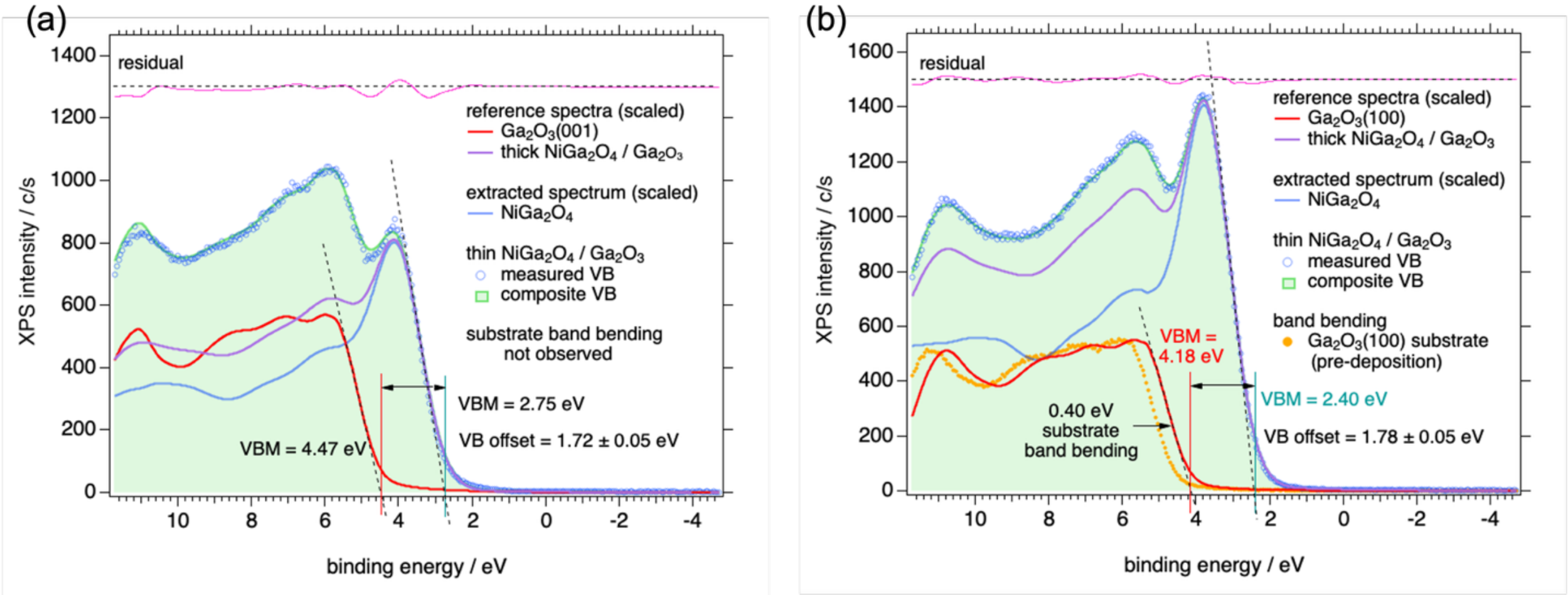

Figure S10: XPS Valence band spectra of thin $NiGa_2O_4$/ β-$Ga_2O_3$ heterostructure for (a) (001) and (b) (100) orientation, showing extrapolated VB offset of ~ 1.7-1.8 eV.

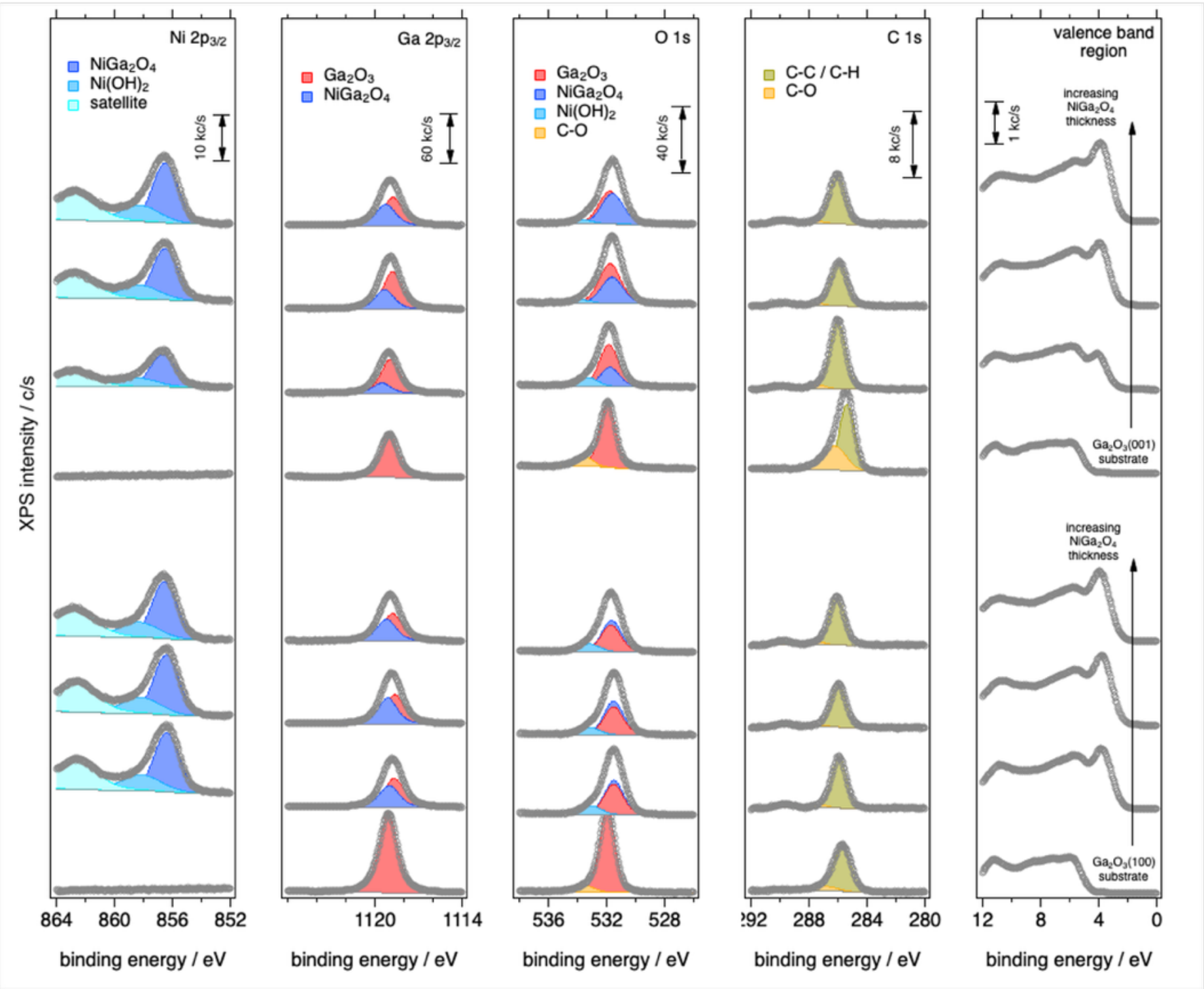

Fig. S11 XPS analysis of $NiGa_2O_4$ on $Ga_2O_3$ substrates.

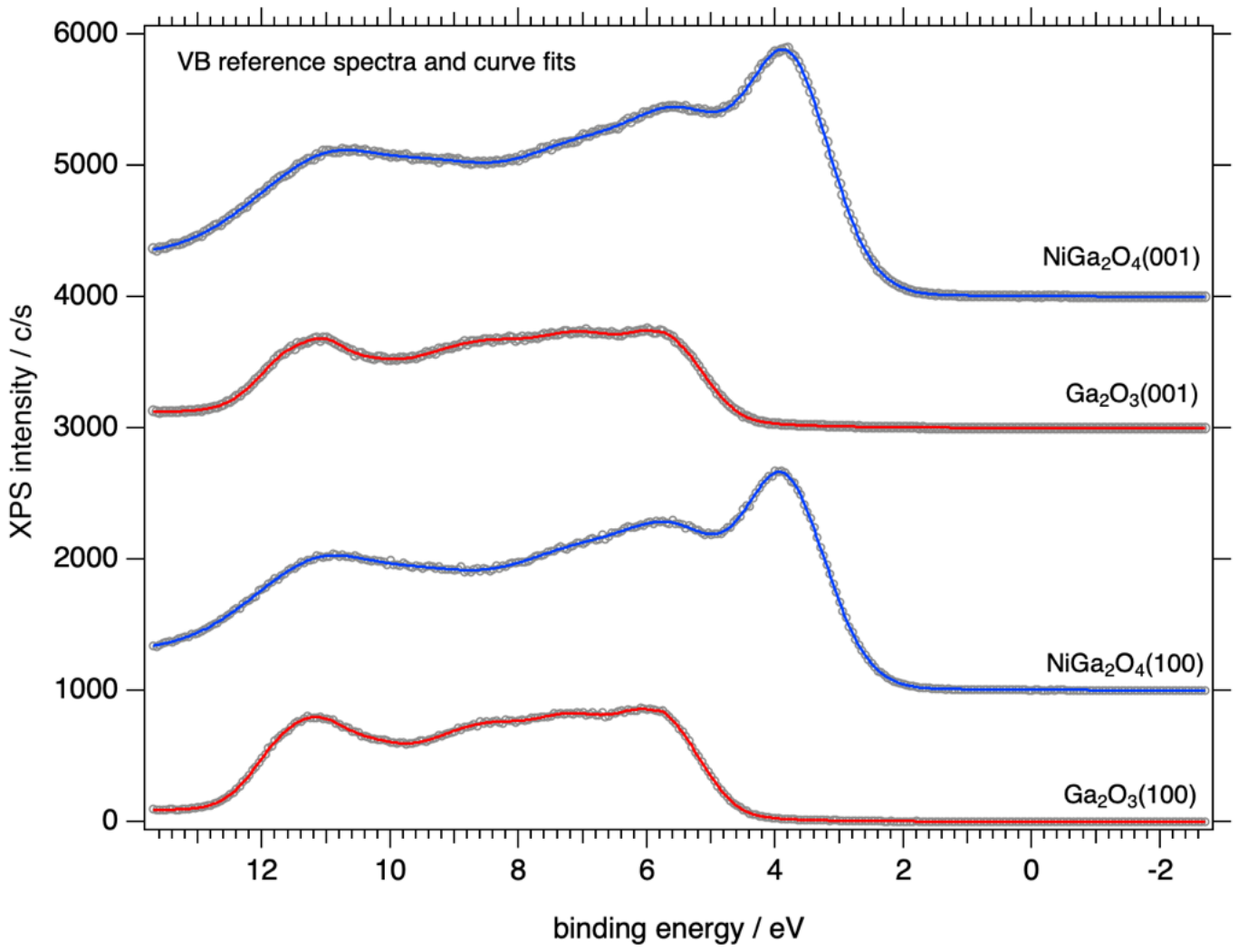

Fig. S12 XPS analysis of $NiGa_2O_4$ on $Ga_2O_3$ substrates

Table S1: Summary of the FWHM values for the different orientation of $NiGa_2O_4$ and the substrate for growths at 700 °C substrate temperature and $p(O_2)$ ~10 mTorr

| Substrate | Layer Orientation | FWHM(°) |
|---|---|---|
| $Al_2O_3$ (00.6) | $NiGa_2O_4$ (111) | 0.059 |
| $Ga_2O_3$ (200) | $NiGa_2O_4$ (200) | 0.48 |
| $Ga_2O_3$ (020) | $NiGa_2O_4$ (220) | 0.82 |

Table S2. Lattice constants (in Angstroms) for $NiGa_2O_4$ *P4₁22* converted into dimensions of the cubic *Fd3m* for easier comparison to measured values of nanocrystals from Ref. Sharma *et al.* For both calculation methods, AFMv1 spin configuration is used. HSE06 agrees closer with experiment.

| | lattice const, a, Å | lattice const, b, Å | lattice const, c, Å | Average lattice constant % difference from experiment |
|---|---|---|---|---|
| Experiment (Sharma et al) | 8.26 | 8.26 | 8.26 | - |
| GGA+U | 8.27 | 8.52 | 8.52 | +2.1% |
| HSE06 (α=0.25) | 8.42 | 8.31 | 8.31 | +1.0% |

Table S3: Device measurement and modeling parameters for the $NiGa_2O4/Ga2O3$ p-n heterojunction devices

| Device | Ideality factor (η) | Turn On voltage(V) | $\Phi_B^{JV}$ (eV) | $R_{on\text{-}sp}$ (mΩ-cm$^2$) | Thermionic Emission Barrier (eV) | Poole-Frenkel Emission Barrier (eV) |
|---|---|---|---|---|---|---|
| $NiGa_2O_4/Ga_2O_3$ | 1.6 | 1.5 | 1.6 | 49 | 1.12 | 1.32 |